\documentclass[twocolumn,superscriptaddress,longbibliography,aps,prl,preprintnumbers]{revtex4-2}

\usepackage[normalem]{ulem}
\usepackage{graphicx}
\usepackage{bm}
\usepackage{color}
\usepackage{epsfig}
\usepackage{amsmath}
\usepackage{amssymb}
\usepackage{wrapfig}

\usepackage{hyperref}
\hypersetup{
	colorlinks=true,      
	urlcolor=blue,
	citecolor=blue,
	linkcolor=blue
}
\footnotetext{These authors contributed equally to this work.}
\begin{document}
	\title{Frequency-selective amplification of nonlinear response in strongly correlated bosons}
	
	\author{Aditya Prakash$^*$}
	\affiliation{National Institute of Science Education and Research, Jatni, Odisha 752050, India}
	\affiliation{Homi Bhabha National Institute, Training School Complex, Anushakti Nagar, Mumbai 400094, India}
	
	\author{Debamalya Dutta$^*$}
	\affiliation{National Institute of Science Education and Research, Jatni, Odisha 752050, India}
	\affiliation{Homi Bhabha National Institute, Training School Complex, Anushakti Nagar, Mumbai 400094, India}
	
	\author{Arko Roy}
	\affiliation{School of Physical Sciences, Indian Institute of Technology Mandi, Mandi-175075 (H.P.), India}
	
	\author{Kush Saha}
	\affiliation{National Institute of Science Education and Research, Jatni, Odisha 752050, India}
	\affiliation{Homi Bhabha National Institute, Training School Complex, Anushakti Nagar, Mumbai 400094, India}

	\begin{abstract}
		
		We present a protocol to generate enhanced non-linear responses of incident pulses in the density wave phase within the extended Bose-Hubbard model using the concept of resonance-induced amplification (RIA). This method enables the selection of an incident pulse frequency to amplify the desired harmonic order. We characterize the enhancement of the non-linear harmonic spectra under various frequencies and field strengths of the incident pulses, and demonstrate that an optimal field strength is necessary to realize our protocol. 
		
	\end{abstract}

	\maketitle
	
	{\em Introduction.-} The strong-field driven dynamics and the phenomenon of high-harmonic generation (HHG) are well studied examples of non-perturbative and nonlinear optical processes \cite{RevModPhys.72.545, Corkum2007}, with the potential to produce attosecond light pulses \cite{Cavalieri2007,RevModPhys.81.163}. It has emerged as a promising platform for the development of ultrafast and short-wavelength coherent light sources \cite{PhysRevA.49.2117}. 
	HHG has been initially studied in atomic and molecular gas systems, in which a characteristic plateau with a cutoff energy is well explained by the three-step model \cite{PhysRevLett.71.1994,Ishika,KO,Agostini_2004, Constant_1999}. Extensive research has demonstrated that HHG in solid-state systems can serve as a valuable tool for investigating the electronic properties of diverse materials \cite{PhysRevLett.115.193603,Ghimire2011}. Strongly correlated neutral cold-atom system have also offered an ideal platform to test the phenomenon of HHG and study the emanating non-linear dynamics associated with strong-field driving.
	Numerous studies \cite{IP1,IP2,IP3,IP4,KS1,KS2,PhysRevA.107.043111} indicate that the non-linear process accompanied by the production of high harmonic orders is indeed a reliable method for probing strongly correlated phases. However, the low efficiency of HHG in these systems limits its applications. Thus, manipulation, enhancement and optimisation of harmonic orders in neutral as well as in electronic systems is the current necessity and the logical next step. Notably, enhancement in HHG has been observed in atomic systems, both experimentally and theoretically \cite{Be,Plummer_2002,Iva,QU,Wang:17}. Recently, there has also been an interest for enhancing HHG in solids \cite{Yoshikawa2019}. That said, a protocol to amplify specific harmonic orders on-demand in neutral systems is presently unknown. Amplifying HHG can lead to more efficient and powerful sources of coherent short-wavelength light \cite{SINGH2024100043,Bruner:18,Atto_pulse}. Few previous works have shown enhancement in the magnitude of higher harmonic orders\cite{QCP,PhysRevA.94.023419}. Shao et al. \cite{QCP} demonstrated that the  spectra is enhanced near QCPs, and suppressed due to flatness of band structures. However, navigating and pinpointing QCPs is itself a challenging task both experimentally and theoretically \cite{Yao,song2023deconfined}.
	
	In this work, we introduce a novel protocol for generating enhanced non-linear responses at specific incident frequencies for the one dimensional (1D) extended Bose Hubbard model (EBHM) within the bosonic density-wave (DW) phase. In particular, we show that the enhancement of the desired harmonic occurs only for a particular choice of frequency, whose integer multiple matches with the energy of the excited states available in the DW phase. Thus, we propose to designate this phenomenon as \textit{resonance-induced amplification} (RIA). In addition to characterizing the {\it analogous} HHG spectra associated with the Haldane insulator (HI), Mott insulator (MI), and DW phases, our investigation aims to address the following key questions: (i) Apart from DW phase why is RIA not feasible in other insulating phases of EBHM, namely MI and HI? (ii) How to select the frequency of the incident pulse in order to amplify the desired harmonic? (iii) 
	How does the amplitude of the incident pulse affect the RIA? 
	To address these fundamental questions, we study 1D EBHM 
	and utilize the well-established zero-temperature phase diagram of the EBHM with model parameters similar to Ref.~\cite{Ent}.

	\begin{figure*}
		\centering
		\includegraphics[width = 1\textwidth]{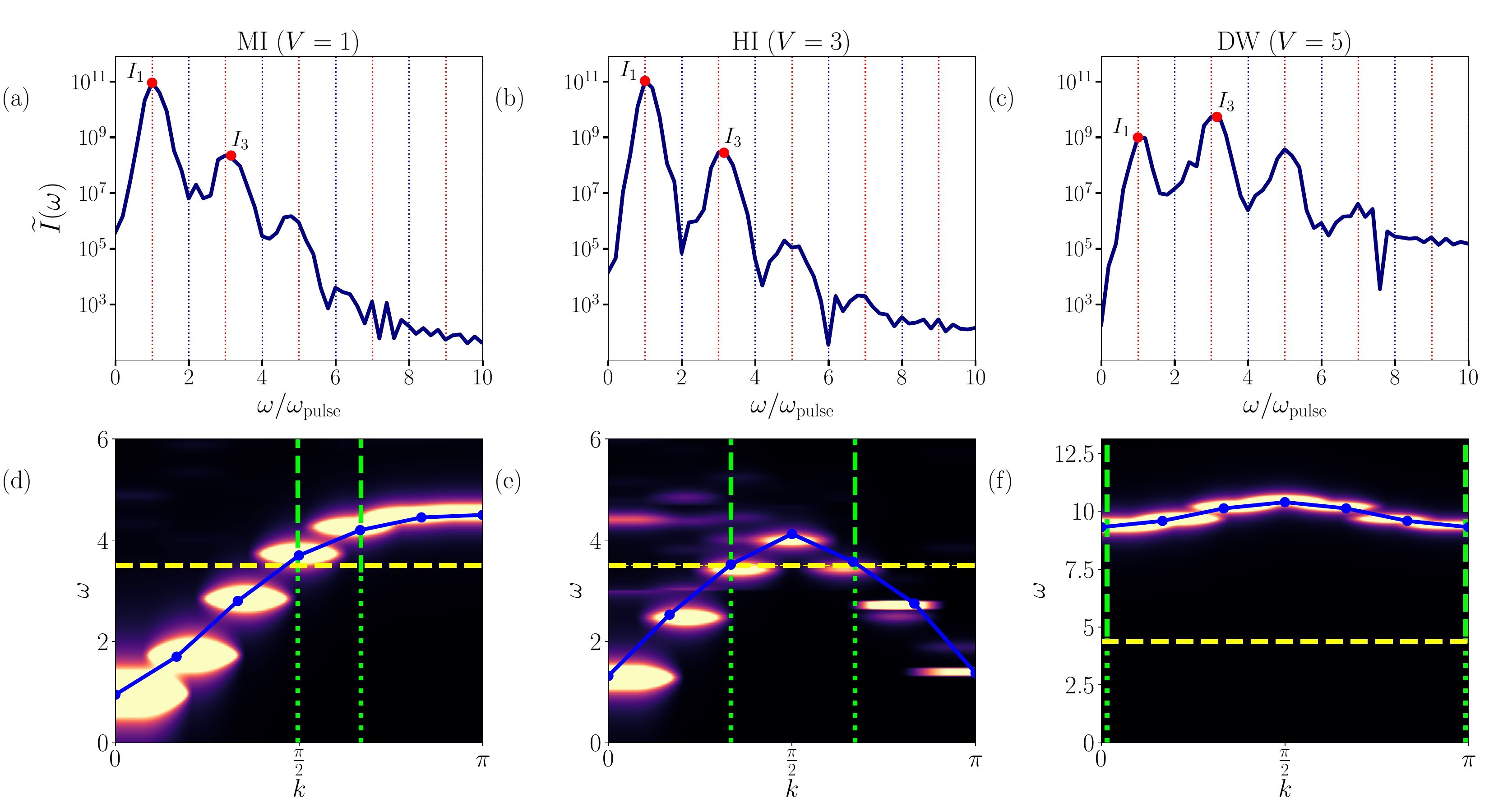}
		\caption{(a)-(c) HHG spectra $\widetilde{I}(\omega)$ 
			(in arbitrary units) at three distinct phases of the extended Bose-Hubbard Model (EBHM), where the maximum boson occupancy is set to $n_{\mathrm{max}} = 2$. These phases correspond to Mott Insulator (MI) with $V=1$, Haldane Insulator (HI) with $V=3$, and Density Wave (DW) with $V=5$, while maintaining the on-site potential at $U=5$. The incident pulse is characterized by frequency $\omega_{\mathrm{pulse}} = 3.5$ and field strength  $A_0 =1$. HHG spectra were computed using the Time-Evolving Block Decimation (TEBD) technique for a system size of $L=50$ with integer filling $\rho = 1$. The red dotted lines here highlight odd harmonics whereas the blue ones indicate even harmonics.
			(d)-(f) The excitation spectrum $S(k,\omega)$ of EBHM. The dark regions denote the absence of any excited states, and the bright regions show the distribution of the available excited states. The yellow dotted line in each of the three plots represent the energy level  $\omega_{\mathrm{pulse}} =3.5$. The green dotted lines show the segment of the momenta $k$ that most dominantly contributes to the HHG spectra. The blue dots represent the intensity maximum of excitation spectrum at the corresponding momenta.}
		\label{Fig1}
	\end{figure*}

	\begin{figure}
		\centering
		\includegraphics[width = \linewidth]{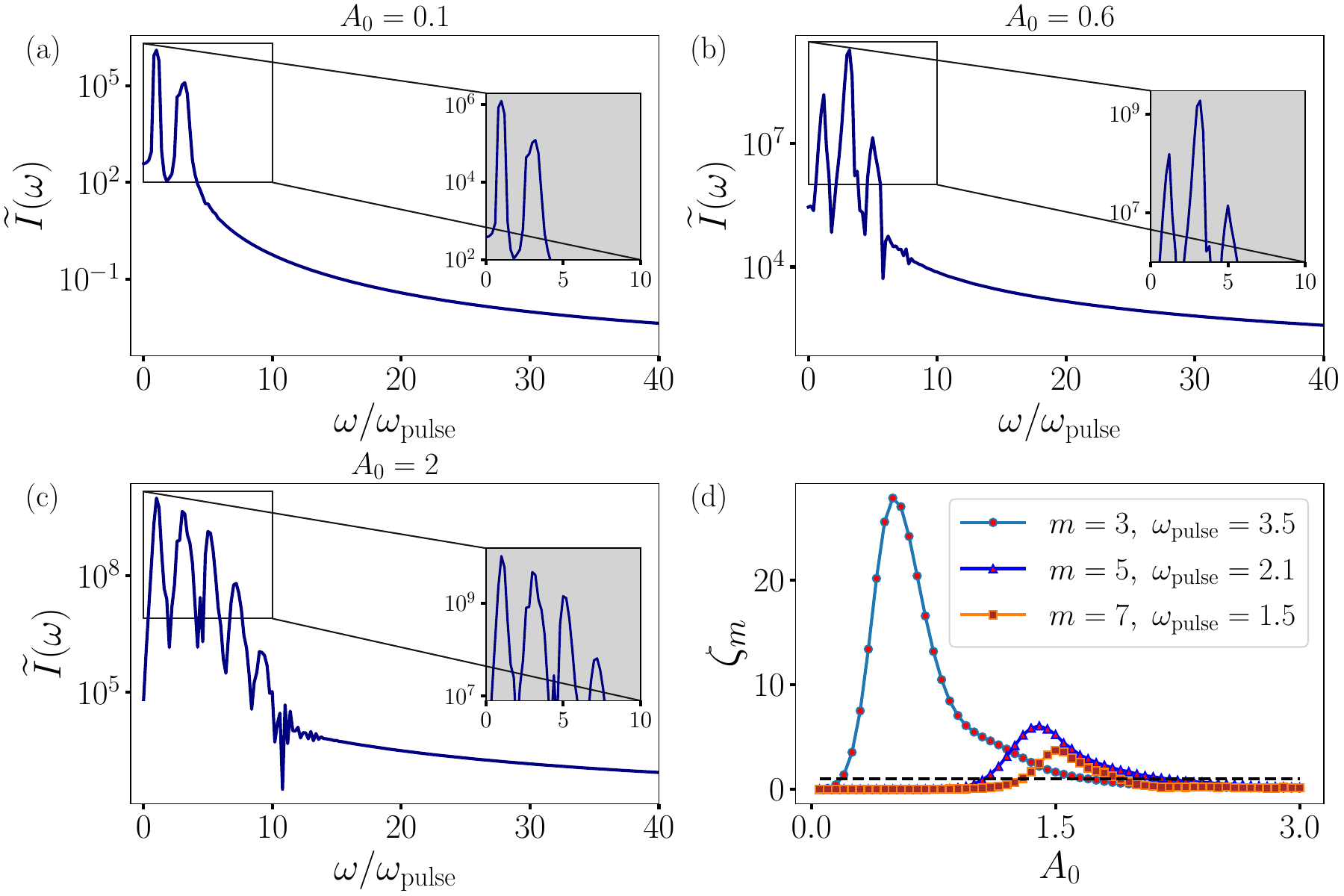}
		\caption{The variation of high-harmonic intensity $\widetilde{I}(\omega)$ 
			(in arbitrary units) across different field strengths in the DW phase ($V=5$) with an input frequency of $\omega_{\mathrm{pulse}} = 3.5$. The zoomed inset in each of the figures helps in comparing the peak of first and third harmonic. Fig.~\ref{Fig5} (a) illustrates the absence of RIA, indicating that at extremely low field strengths, RIA is not achievable. Fig.~\ref{Fig5} (b) demonstrates RIA, characterized by a significant  enhancement of the third harmonic as compared to the first harmonic. Fig.~\ref{Fig5} (c) highlights that RIA vanishes at extremely intense field strengths.
			Fig.~\ref{Fig5} (d) illustrates the RIA of various harmonics for different
			incident pulse frequencies and varying field strength. Specifically, an incident
			pulse frequency of $\omega_{\mathrm{pulse}} = 3.5$ enhances the third harmonic,
			while $\omega_{\mathrm{pulse}} = 2.1$ boosts the fifth harmonic. Similarly,
			$\omega_{\mathrm{pulse}} = 1.5$ results in the amplification of the seventh
			harmonic, however for $U=1, V=4$ (see Supplementary Material). }
		\label{Fig5}
	\end{figure}

	{\em Model.-} The 1D EBHM with on-site and nearest-neighbour repulsive interactions is given by:
	\begin{eqnarray}
		H &=& -J \sum_j \left( b_j^\dagger b_{j+1} + H.c. \right) + \frac{U}{2} \sum_j n_j (n_j-1)\nonumber\\ &+&V \sum_j n_j n_{j+1}, 
	\end{eqnarray}
	where $b^\dagger_j, \, b_j$ are creation and annihilation operators of bosons on site $j$.
	$J$ denotes the hopping strength, while $U$ and $V$ are the repulsive on-site and the nearest-neighbour interaction strengths. With mean filling ${\rho} = N/L =1$ and maximum of two bosons per site ($n_{\mathrm{max}} =2$), we confirm that the phases in the  phase diagram~\cite{Ent} remain valid for system size $L=50$, enabling us to conduct HHG simulations in these phases at this system size. The process of HHG in electronic systems involves the electric field $E(t)$ interacting with electrons through the time-varying vector potential $A(t)$. In contrast, an analogous HHG in bosonic neutral atoms is synthetically induced by the coupling between $A(t)$ and neutral atoms. Experimentally, it has been demonstrated that time variation of the vector potential produces a synthetic electric field ($E(t)=-\partial_t A(t)$) that acts on the neutral particles as does a real electric field on charged particles \cite{synthetic}. As the light couples to the Hamiltonian, the tunnelling term $J$ becomes complex as it acquires the Peierls phase \cite{Peierls1933}. The effective time-dependent tunneling in the velocity gauge takes the form given by $J(t)\equiv Je^{{i} \Phi(t)}$ with $\Phi(t)=qA(t)a/\hbar$, where $a$ represents the lattice parameter, and $q$ signifies the effective charge of the boson \cite{synthetic}. In the present study, we employ a $n=5$-cycle $\sin^2$ time-varying potential in the form of a pulse $A(t)=A_0\sin^2{(\omega_{\mathrm{pulse}} t/2n)}\sin{(\omega_{\mathrm{pulse}} t)}$ where $\omega_{\mathrm{pulse}}$ denotes the oscillation frequency and $n$ denotes the number of cycles. In the rest of our investigation, we express $A_0$ in dimensionless units. We begin by calculating the ground state of the wave function using the Density Matrix Renormalization Group (DMRG). This ground state is then utilized to study its evolution under the influence of a vector potential, employing the time-evolving block decimation (TEBD) technique. We use the exact diagonalisation (ED) method to calculate the excitation spectrum. All these numerical techniques are implemented using the open-source Python-based libraries Tenpy \cite{tenpy}, Quspin \cite{Quspin} and C++ library ITensor\cite{itensor}.
	We set the bond dimensions to $\chi$ = 800, applying the fourth-order Trotter decomposition with an optimum temporal step size. We set the energy scale using $J=1$ and work in units where $\hbar= a =  q = 1$. 
	For typical values with $J\sim {\mathrm{meV}}$ and $a\sim \text{\AA}$, the bandgap $\omega$ lies within the experimentally feasible THz range, and the electric field in $\mathrm{MVm^{-1}}$ range. To investigate the effect of synthetic electric field, we first evaluate current operator defined as 
	\begin{align}
		\mathcal{J}(t)= -\frac{\partial {H}}{\partial A}=J\sum_{j=1} \left(ie^{iA(t)}b_{j}^{\dagger}b_{j+1}+{\mathrm{H.c.}}\right).
	\end{align}
	The response is proportional to the dipole acceleration $I(t)\propto\langle\mathcal{\dot{J}}(t)\rangle$ \cite{Gaarde_2008}, where $\langle\mathcal{J}(t)\rangle=\langle\psi_0(t)|\mathcal{J}(t)|\psi_0(t)\rangle$, and $|\psi_0(t)\rangle$ represents the time evolved ground state of EBHM calculated numerically. The intensity $\widetilde{I}(\omega)=|{\mathrm{FT}}[I(t)]|^2$, where ${\mathrm{FT}}[..]$ represents the Fourier transform. The distribution of the higher harmonic orders for three distinct insulating phases of the EBHM (MI ($V=1$), HI ($V=3$) and DW($V=5$)) are shown by Fig.~\ref{Fig1}(a)-(c).  In order to analyse the HHG results and understand the distribution of excited states, we calculate the dynamic structure factor, which carries information of the excitation spectrum of the system \cite{Sturm,David,Bragg1}. The dynamic structure factor 
	$S(k, \omega) = \sum_{m} |\langle \psi_m | n_k | \psi_0 \rangle|^2 
	\delta(\omega - \omega_m)$, with $|\psi_m\rangle$ and $\omega_m$ being the $m^{th}$ eigen state and corresponding eigen value respectively, can be expressed as

	\begin{equation}
		S(k,\omega)= \frac{1}{2\pi} \sum_{j\,j'}\int e^{-i\omega t} e^{-i{k}(j-j')} \langle \hat{n}_j(t) \hat{n}_{j'}(0) \rangle\,dt\label{eq:ds},
	\end{equation}
	
	\noindent
	where $\langle n_j(t) n_{j'}(0) \rangle$ denotes the expectation value of 
	density-density correlations with respect to the ground state of EBHM. Using Eq.~(\ref{eq:ds}), Figs.~\ref{Fig1}(d-f) display the excitation spectrum $S(k,\omega)$ of EBHM at $U=5$ and different values of $V$. Fig.~\ref{Fig1}(d) shows the excitation spectra for MI ($V=1$), where the spectral weight can be seen to get flattened in the region $k> 2.1$, around $\omega\simeq 5$. The excitation spectra changes significantly when we enter the HI ($V=3$) phase in Fig.~\ref{Fig1}(e), where we find a sine like dispersion. Fig.~\ref{Fig1}(f) shows the bosonic DW ($V=5$) phase, whose excitation spectra is relatively special as it shows an almost flat behaviour. This implies that all the momenta $k$ require the same amount of energy to become excited. Additionally, the excited states available in the DW phase are at higher energies compared to those in other insulating phases (MI and HI), as evidenced by the $\omega$ values shown in Fig.~\ref{Fig1} (d)-(f). Our excitation spectrum align with the findings of the EBHM detailed in Ref.~\cite{Ent}. Furthermore, since the EBHM can be mapped to a spin-1 anisotropic Heisenberg chain \cite{OGolinelli_1993,PhysRevB.75.085106,PhysRevLett.97.260401,PhysRevLett.89.250404}, our results for the excitation spectrum match those found for the spin model, as shown in Ref.~\cite{SpinEnt}. In the upcoming discussion, we will recognize how these features aid DW phase in demonstrating RIA.

	{\em RIA.-} Fig.~\ref{Fig1}(c) shows that the third harmonic is anomalously enhanced,
	with its amplification being even greater than that of the first harmonic. This is in
	stark contrast to Figs.~\ref{Fig1}(a) and (b). Where it does not show any harmonic
	with greater intensity than the first harmonic. The enhancement happens as the third
	integer multiple of the incident pulse frequency ($\omega_{\mathrm{pulse}} = 3.5$)
	nearly matches the energy of the excitation spectrum of the DW phase
	($\omega_{\mathrm{DW}} = 10.5$), as illustrated in Fig.~\ref{Fig1}(f)). Due to the
	near match of the energies (energy of the third multiple of incident pulse frequency
	and energy of the excitation spectrum), the system prefers this transition, which
	results in the amplification of the corresponding (third) harmonic. This occurs only
	in the DW phase because its excitation spectrum is relatively flat. This means that
	the system has a fixed threshold energy for all momenta to get excited, and unlike
	other phases, all momenta de-excite emitting the same harmonic frequency.
	Additionally, the excited states in the DW phase are well above the incident pulse
	frequency, allowing all momenta to contribute to HHG. In other phases, only a few
	momenta have energies above the incident pulse frequency, so different momenta
	contribute different harmonics. Since this enhancement depends on the choice of the
	frequency and occurs only when the multiple of the incident pulse matches the energy
	of the available excited states, we call this \textit{resonance induced
		amplification} (RIA). The anomalous enhancement disappears when the
	incident pulse frequency does not match the energy of the excited states.
	For instance, $\omega_{\rm pulse} = 6.5$ and a field strength of $A_0=1$
	does not facilitate enhancement of the third harmonic. Instead, as usual, the first 
	harmonic is the most intense signal (see Supplementary Material).
	
	{\em Effect of amplitude.-} Fig.~\ref{Fig5} shows the HHG spectra in the DW phase for an incident pulse frequency $\omega_{\mathrm{pulse}} =3.5$ at three different amplitudes.
	In Fig.~\ref{Fig5}(a), for a field strength of $A_0 = 0.1$, even though the frequency multiple is resonant with the energy of the excited states, the intensity of the third harmonic does not surpass the intensity of the first harmonic. At very low field strength, the system is still close to the linear response regime, inhibiting RIA.
	However, in Fig.~\ref{Fig5}b, with a slightly increased field strength, RIA is observed, and the intensity of the third harmonic surpasses the first. Interestingly, in Fig.~\ref{Fig5}(c), with the field strength set to $A_0 = 2$, the third harmonic is no longer the most intense signal, and the usual trend resumes, with the first harmonic being the most intense. The overall variation of $\zeta_m = {\widetilde{I}}_m/{\widetilde{I}}_1$ with the field strength $A_0$ is encapsulated in Fig.~\ref{Fig5} (d) with
	${\widetilde{I}}_1$ and ${\widetilde{I}}_m$ being the intensities of the first and $m^{th}$ harmonic. This reveals that RIA cannot occur at very low or very high field strengths. 
	
\begin{figure}
		\centering
		\includegraphics[width = \linewidth]{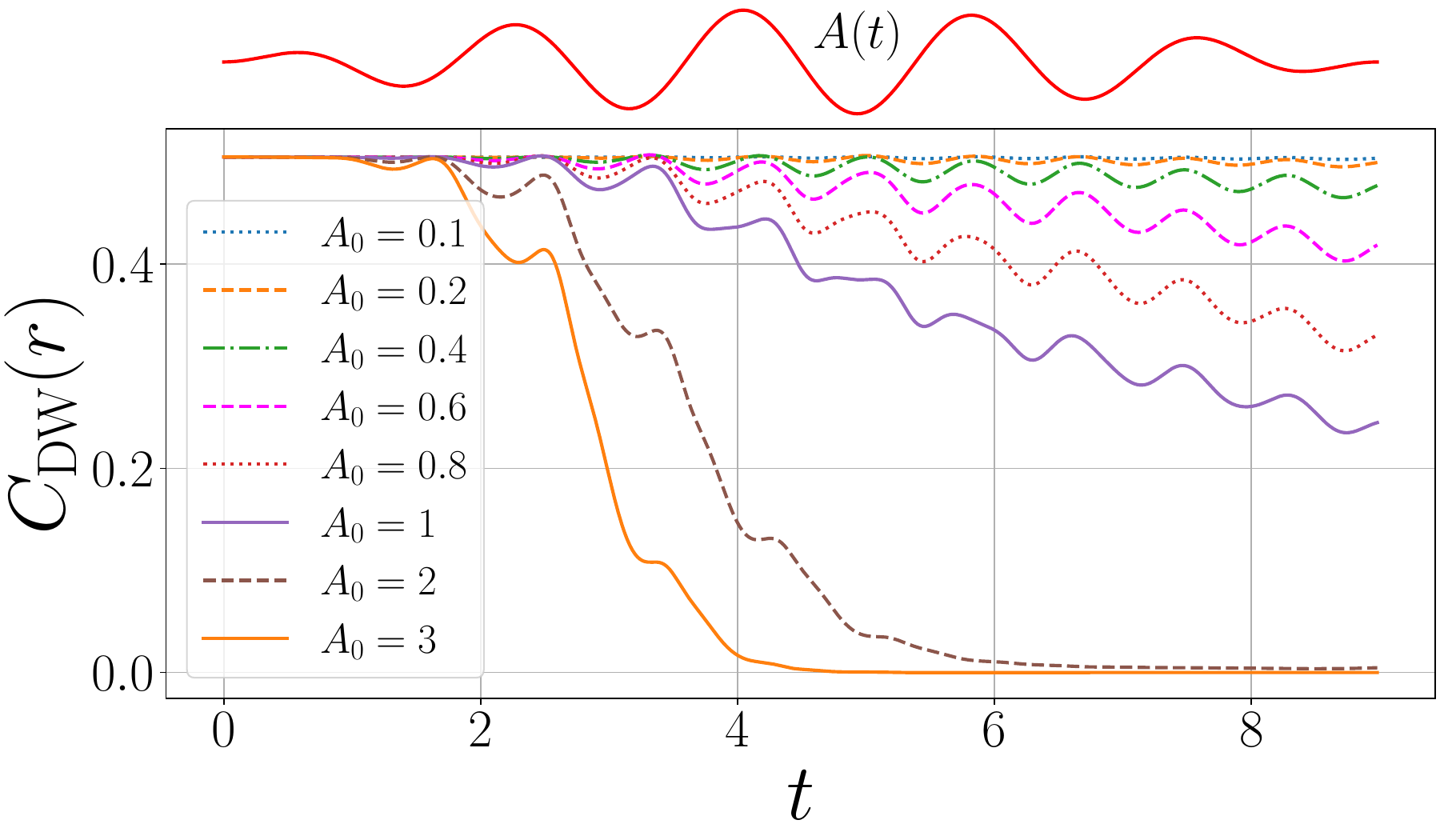}
		\caption{The temporal evolution of the order parameter of the Density Wave (DW) phase ($C_{\mathrm{DW}}(r) = (-1)^r \langle \delta n_j \delta n_{j+r} \rangle$) is investigated, emphasizing the impact of an incident pulse with a frequency of $\omega_{\mathrm{pulse}} = 3.5$ and at various field strengths. Notably, field strengths equal to or exceeding $A = 2$ are found to entirely disrupt the phase, consequently halting RIA as well.}
		\label{Fig6}
	\end{figure}

	This can be understood by examining the variation of the DW phase order parameter, given by $C_{\mathrm{DW}}(r)   =  (-1)^r \langle \delta n_j \delta n_{j+r} \rangle$
	with $\delta n_j = n_j  - \rho$. To remove  boundary effects, we computed the average expectation values of all the two-point correlators between bosons in the middle part of the chain, discarding the outer $L/4$ sites on both sides ($j = L/4, r = L/2$). Fig.~\ref{Fig6} shows the variation of $C_{\mathrm{DW}}(r)$ with time under the influence of the synthetic electric field. We notice that for very high field strengths like $A_0 = 2$ or $3$, the order parameter $C_{\mathrm{DW}}(r)$ vanishes completely within the period of the light pulse. This indicates that at very high field strengths, the DW phase is destroyed, eliminating RIA. This also shows that RIA is a signature property of the DW phase.
	To conclude, for successful RIA generation, the amplitude must be carefully optimized: it should be high enough to induce nonlinear effects but not so high that it disrupts the phase entirely within the pulse duration.

	{\em Frequency selection.-} 
	To further elaborate the manifestation of RIA at other input frequencies we refer to Fig.~\ref{Fig5}(d). We emphasize that when the multiples of the incident pulse frequency matches with the energy of the excited states for the DW phase, it gives rise to the resonance conditions. That is, $\omega_{\mathrm{pulse}} = 3.5, 2.1$ and $1.5$ results in amplification of the third, fifth, and seventh harmonic, albeit at an optimum field strength. Fig.~\ref{Fig5} (d) shows that at very high field strengths, there is no RIA because the DW phase is destroyed as evident from Fig.~\ref{Fig6}. Additionally, we observe that the peaks shift to the right, indicating that higher field strength is needed to manifest RIA in higher harmonics. The peak of $\zeta$ decreases as the frequency lowers, making it harder to enhance higher-order harmonics. Therefore, it is challenging to observe significant RIA at higher harmonic orders. As higher field strength is required to realize RIA for enhancement of higher harmonics, the DW phase needs to be more stable and robust under the influence of light pulse. To achieve RIA for higher harmonics, it is essential to find robust DW phase regions, which do not decay on shining stronger light fields (see Figs.~\ref{Fig5},\ref{Fig6}, and also Supplementary Material).

	{\em Conclusion.-} Our study introduces the concept of RIA, which shows the potential to enhance the output signals (with respect to the first harmonic) of an analogous HHG process in neutral bosonic systems. Further, our protocol provides a clear method for selecting incident pulse frequencies to control harmonic amplification. We have examined the enhancement in higher harmonic order spectra and analyzed its behaviour under different frequencies and field strengths. Our results highlight the unique properties of the excitation spectrum in the DW phase, which supports the occurrence of RIA. We demonstrated that RIA is a characteristic property of the DW phase, where the near flat excitation spectrum allows for a fixed energy threshold to excite all momenta simultaneously. This makes the system prefer transitions where the frequency multiple matches the energy of the excited states, amplifying specific signals. By choosing the incident pulse frequency to match these energies, we can amplify the desired harmonic, allowing precise control over which harmonic is amplified. We also found that the field strength of the incident pulse is crucial for RIA. To effectively generate RIA, the field strength must be strong enough to produce nonlinear effects but not so strong that it completely destroys the DW phase. This work not only contributes to the efficient generation of HHG, but also introduces a novel technique for amplifying specific harmonics in the HHG spectra, paving the way for more powerful and coherent short-wavelength light sources and intense attosecond pulses without the need for filtering in experiments. With the experimental realization of EBHM in different settings and proposal to  access the emission spectrum of HHG in cold-atom simulators~\cite{weckesser2024realizationrydbergdressedextendedbose,Lagoin2022,Weld}, we believe our protocol can be tested in the near future.

	\textit{Acknowledgements: } AP would like to thank Bharath Hebbe Madhusudhana for the valuable numerical training and insight. AP would also like to thank Tapan Mishra and Ashirbad Padhan for useful discussions. DD and AP acknowledge the Virgo cluster, where most of the numerical calculations were performed. AR acknowledges the support of the Science and Engineering Research Board (SERB), Department of Science and Technology, Government of India, under the project SRG/2022/000057 and IIT Mandi seed-grant funds under the project IITM/SG/AR/87. KS and DD acknowledge financial support from the Department of Atomic Energy (DAE), Govt. of India, through the project Basic Research in Physical and Multidisciplinary Sciences via RIN4001. KS acknowledges funding from the Science and Engineering Research Board (SERB) under SERB-MATRICS Grant No. MTR/2023/000743. 
	
	\onecolumngrid
	\section*{Supplementary Material: Frequency-selective amplification of nonlinear response in strongly correlated bosons}

	This Supplementary Material provides further details on the harmonic spectra in distinct phases of EBHM for a range of applied field and the generation of resonance-induced amplification (RIA) for various amplitudes and incident frequencies in the DW phase. 
	
	\section{Harmonic spectra in three distinct phases of extended Bose Hubbard Model}
\begin{figure*}[b]
		\centering
		\includegraphics[width =0.75\linewidth]{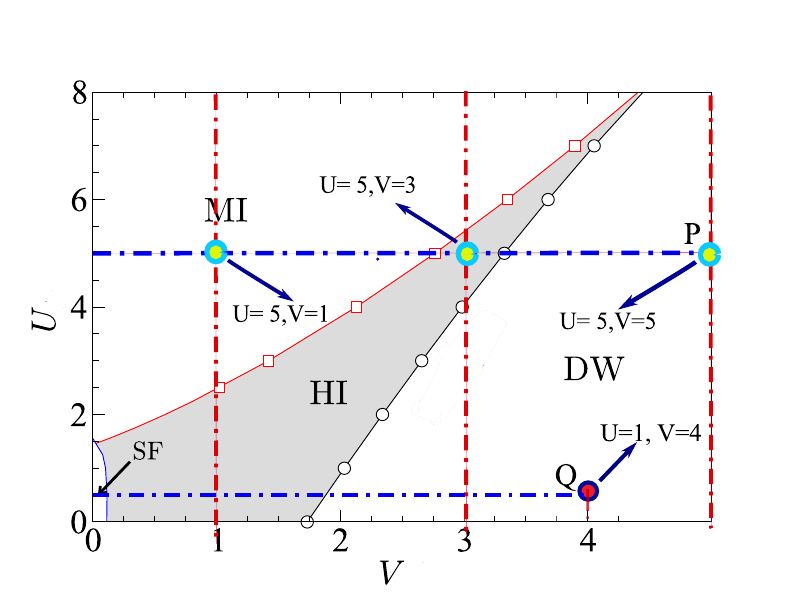}
		\caption{ Phase diagram of EBHM by Ejima et al.\cite{Ent} including points that were used in our HHG spectroscopy. It depicts the phase diagram of EBHM for mean filling $\rho =1$ and $n_{max} =2$. The points marked in circles, with values $U=5,V=1$, $U=5,V=3$, and $U=5,V=5$, represent the MI, HI and DW phases, respectively. We coupled light to these points and study harmonics in the particle current.}
		\label{phase}
	\end{figure*}

	The enhancement of a particular harmonic order under resonant condition is a distinct feature of DW phase as opposed to the MI and HI phases which do not exhibit such features for any range of applied field. To demonstrate this, we first cite the phase diagram of EBHM in Fig.~\ref{phase} taken from Ref.~\cite{Ent}, highlighting three points in MI ($U=5,\,V=1$), HI ($U=5,\,V=3$) and DW ($U=5,\,V=5$) phases for our calculations. 
	
\begin{figure}[t]
		\centering
		\includegraphics[width  = \linewidth]{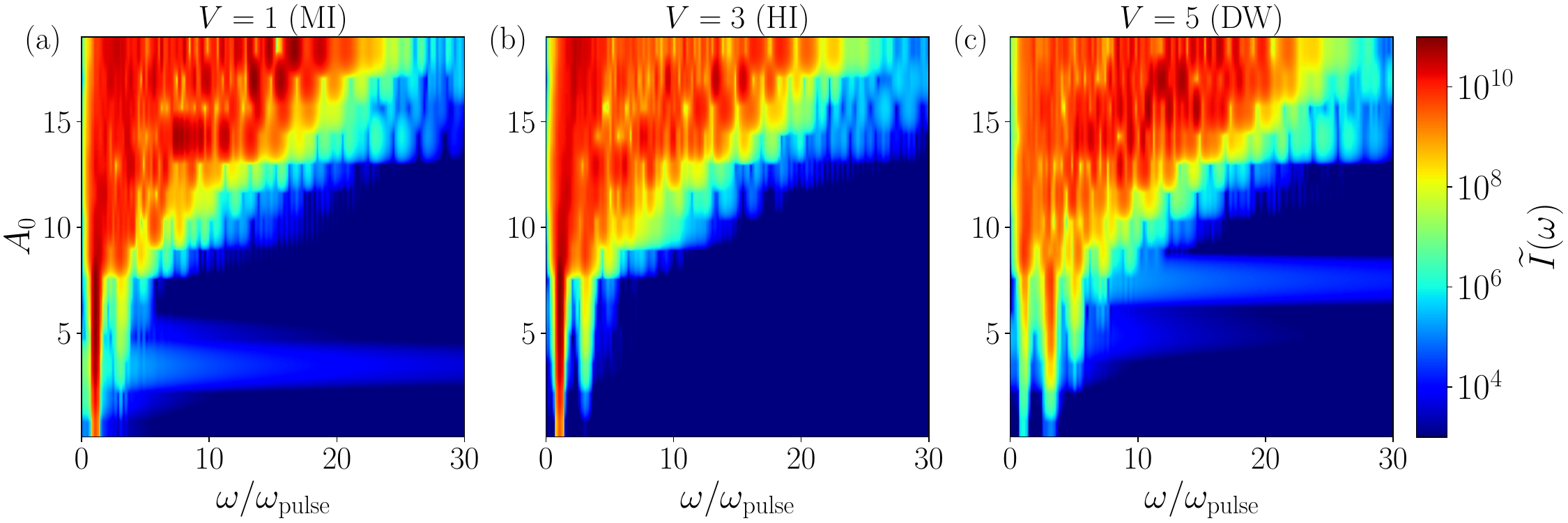}
		\caption{False color-coded images show the intensity variations of HHG spectra at different harmonics for varying field strengths, with a pulse frequency of $\omega_{\mathrm{pulse}}=3.5$. In both the MI (a) and HI (b) phases, the first harmonic is the most intense across different field strengths. However, in the DW phase (c), the third harmonic becomes the most intense, indicating the presence of RIA. This characteristics is intrinsic to the system in DW phase.}
		\label{fig:color_plot_RIA}
	\end{figure}

	Fig.~\ref{fig:color_plot_RIA} illustrates intensity spectra for different field strengths $A_0$ with pulse frequency $\omega_{\mathrm{pulse}}=3.5$. In the MI (Fig.~\ref{fig:color_plot_RIA}(a)) and HI (Fig.~\ref{fig:color_plot_RIA} (b)) phases, the first harmonic is consistently the most intense up to the threshold field strength. However, in the DW (Fig.~\ref{fig:color_plot_RIA}(c)) phase, the third harmonic is the most intense as the energy of the excitation gap ($\omega_{\mathrm{DW}}=10.5$) matches the third harmonic of the incident pulse frequency ($\omega_{\mathrm{pulse}}=3.5$). This result indicates that RIA is an intrinsic property of the DW phase and does not manifest in other phases, regardless of adjustments to the frequency and field strength of the incident pulse.

	\begin{figure}[b]
		\centering
		\includegraphics[width=\linewidth]{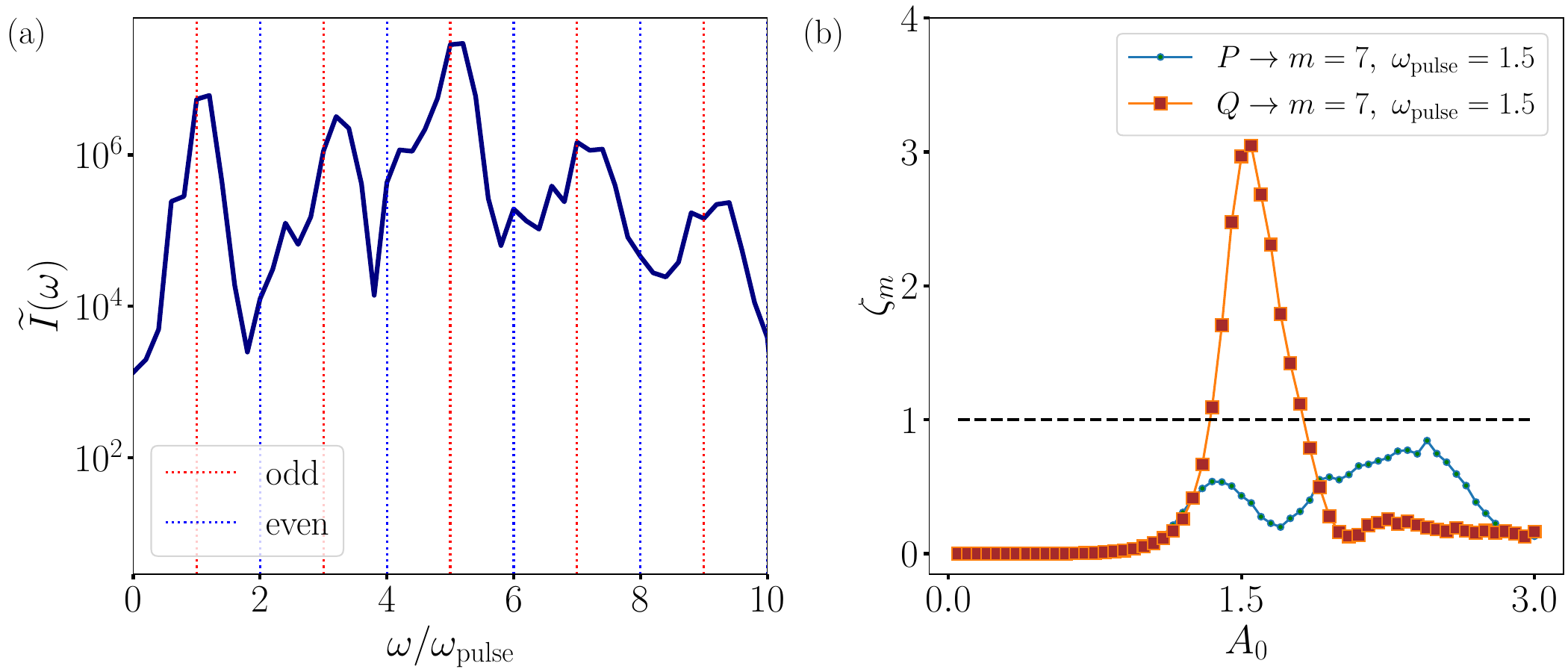}
		\caption{(a) Harmonic spectra in the DW phase ($V = 5$, $U = 5$), using an incident pulse frequency of $\omega_{\mathrm{pulse}} = 2.1$ and a field strength of $A_0 = 1.5$. It highlights the anomalous enhancement of the $5^{th}$ harmonic relative to the $1^{st}$ harmonic. (b)  Variation of the parameter $\zeta_m=\widetilde I_{7}/\widetilde I_{1}$ as a function of filed strength $A_0$ for an incident pulse frequency $\omega_{\mathrm{pulse}} = 1.5$. The horizontal black dotted line represents $\zeta_m = 1$. It is evident that the anomalous enhancement of the $7^{th}$ harmonic relative to the $1^{st}$ harmonic is possible at point $Q$ in contrast to the DW phase point $P$ on the phase diagram.}
		\label{DWS}
	\end{figure}
	
	\section{RIA at different frequencies and field strengths}
	As pointed in the main text, the field-induced enhancement of a particular harmonic order in the DW phase is attributed to the energy match between excitation gap and multiple of incident frequency. For enhancement of higher harmonics, higher field strengths are required, provided that the DW phase is robust and stable. To investigate this,
	\begin{wrapfigure}{r}{0.5\textwidth}
		\begin{center}
			\includegraphics[width=\linewidth]{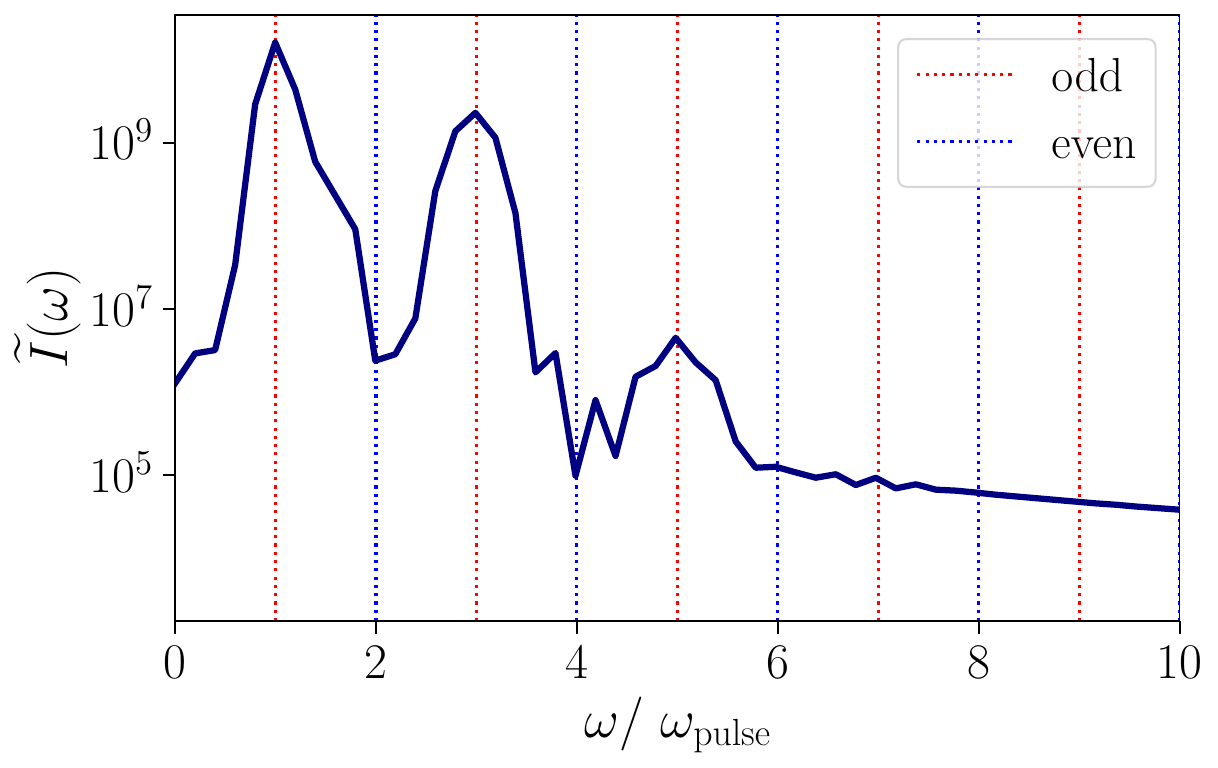}
		\end{center}
		\caption{High Harmonic Generation (HHG) spectra in the DW phase ($V = 5$, $U = 5$), using an incident pulse frequency of $\omega_{\mathrm{pulse}} = 6.5$ and a field strength of $A_0 = 1$. This demonstrates that RIA is unattainable at non-resonant frequencies, as none of the harmonics exhibit anomalous enhancement. Instead, we observe the traditional HHG spectra, where the first harmonic is the most intense signal.}
		\label{fig:HHG_6.5}
	\end{wrapfigure}
	we consider pulse frequency to be $\omega_{\mathrm{pulse}}=2.1$ in the DW phase. The RIA is then expected to be present for $5^{th}$ harmonic as the excitation gap is $\omega_{\mathrm{DW}}\sim 10.5$. Fig.~\ref{DWS}(a) corroborates this as the $5^{th}$ harmonic is dominant but at $A_0=1.5$ as compared to the 3rd harmonic at a relatively lower field (shown in the main text). To tune the $7^{th}$ harmonic, we set $\omega_{\mathrm{pulse}}=1.5$, however, this requires a stronger field that destroys the DW phase at $U=5$ and $V=5$. Interestingly, the DW phase at $U=1$ and $V=4$ as indicated by Q in Fig.~\ref{phase} is found to support enhancement of the $7^{th}$ harmonic at a field strength $A_0=1.5$, comparable to the field strength applied to generate the $5^{th}$ harmonic RIA at the point $P$. As point $Q$ reaches its threshold field amplitude to generate RIA in $7^{th}$ harmonic at lower field amplitude compared to point $P$, the DW order parameter does not completely decay in point $Q$ for such field strength, and we got RIA at the $7^{th}$ harmonic at point $Q$. This is evident from Fig.~\ref{DWS}(b).
	
	To this end, we show that the integer multiple of incident frequency away from the excitation gap does not lead to RIA. Hence  any particular harmonic cannot be enhanced by tuning $A_0$. In doing so, we take incident frequency $\omega_{\mathrm{pulse}}=6.5$. Since the energy gap of the excited states $\omega_{\mathrm{DW}}\sim 10.5$ does not match with any integer multiple of the $\omega_{\mathrm{pulse}}$, the RIA is absent. Accordingly, the first harmonic is dominant and the applied field cannot be used to enhance any particular higher order harmonics as evident from 
	Figure ~\ref{fig:HHG_6.5}(a). 
	
	\vspace{0.5cm}
	
	\twocolumngrid
	
	\bibliography{bibfile}

\begin{thebibliography}{50}%
\makeatletter
\providecommand \@ifxundefined [1]{%
 \@ifx{#1\undefined}
}%
\providecommand \@ifnum [1]{%
 \ifnum #1\expandafter \@firstoftwo
 \else \expandafter \@secondoftwo
 \fi
}%
\providecommand \@ifx [1]{%
 \ifx #1\expandafter \@firstoftwo
 \else \expandafter \@secondoftwo
 \fi
}%
\providecommand \natexlab [1]{#1}%
\providecommand \enquote  [1]{``#1''}%
\providecommand \bibnamefont  [1]{#1}%
\providecommand \bibfnamefont [1]{#1}%
\providecommand \citenamefont [1]{#1}%
\providecommand \href@noop [0]{\@secondoftwo}%
\providecommand \href [0]{\begingroup \@sanitize@url \@href}%
\providecommand \@href[1]{\@@startlink{#1}\@@href}%
\providecommand \@@href[1]{\endgroup#1\@@endlink}%
\providecommand \@sanitize@url [0]{\catcode `\\12\catcode `\$12\catcode
  `\&12\catcode `\#12\catcode `\^12\catcode `\_12\catcode `\%12\relax}%
\providecommand \@@startlink[1]{}%
\providecommand \@@endlink[0]{}%
\providecommand \url  [0]{\begingroup\@sanitize@url \@url }%
\providecommand \@url [1]{\endgroup\@href {#1}{\urlprefix }}%
\providecommand \urlprefix  [0]{URL }%
\providecommand \Eprint [0]{\href }%
\providecommand \doibase [0]{https://doi.org/}%
\providecommand \selectlanguage [0]{\@gobble}%
\providecommand \bibinfo  [0]{\@secondoftwo}%
\providecommand \bibfield  [0]{\@secondoftwo}%
\providecommand \translation [1]{[#1]}%
\providecommand \BibitemOpen [0]{}%
\providecommand \bibitemStop [0]{}%
\providecommand \bibitemNoStop [0]{.\EOS\space}%
\providecommand \EOS [0]{\spacefactor3000\relax}%
\providecommand \BibitemShut  [1]{\csname bibitem#1\endcsname}%
\let\auto@bib@innerbib\@empty
\bibitem [{\citenamefont {Brabec}\ and\ \citenamefont
  {Krausz}(2000)}]{RevModPhys.72.545}%
  \BibitemOpen
  \bibfield  {author} {\bibinfo {author} {\bibfnamefont {T.}~\bibnamefont
  {Brabec}}\ and\ \bibinfo {author} {\bibfnamefont {F.}~\bibnamefont
  {Krausz}},\ }\bibfield  {title} {\bibinfo {title} {Intense few-cycle laser
  fields: Frontiers of nonlinear optics},\ }\href
  {https://doi.org/10.1103/RevModPhys.72.545} {\bibfield  {journal} {\bibinfo
  {journal} {Rev. Mod. Phys.}\ }\textbf {\bibinfo {volume} {72}},\ \bibinfo
  {pages} {545} (\bibinfo {year} {2000})}\BibitemShut {NoStop}%
\bibitem [{\citenamefont {Corkum}\ and\ \citenamefont
  {Krausz}(2007)}]{Corkum2007}%
  \BibitemOpen
  \bibfield  {author} {\bibinfo {author} {\bibfnamefont {P.}~\bibnamefont
  {Corkum}}\ and\ \bibinfo {author} {\bibfnamefont {F.}~\bibnamefont
  {Krausz}},\ }\bibfield  {title} {\bibinfo {title} {Attosecond science},\
  }\href {https://doi.org/10.1038/nphys620} {\bibfield  {journal} {\bibinfo
  {journal} {Nature Phys}\ }\textbf {\bibinfo {volume} {3}},\ \bibinfo {pages}
  {381} (\bibinfo {year} {2007})}\BibitemShut {NoStop}%
\bibitem [{\citenamefont {Cavalieri}\ \emph {et~al.}(2007)\citenamefont
  {Cavalieri}, \citenamefont {M{\"u}ller}, \citenamefont {Uphues},
  \citenamefont {Thiele}, \citenamefont {Schultze}, \citenamefont {Sansone},
  \citenamefont {M{\"u}cke}, \citenamefont {Nolte}, \citenamefont {Stamm},
  \citenamefont {Pr{\"u}mper}, \citenamefont {Rossbach}, \citenamefont
  {Schr{\"o}ter}, \citenamefont {Popmintchev}, \citenamefont {H{\"a}drich},
  \citenamefont {Kling}, \citenamefont {Locher}, \citenamefont {Schreiber},
  \citenamefont {Pietsch}, \citenamefont {Azoury}, \citenamefont {Fischer},
  \citenamefont {Monmayrant}, \citenamefont {Kienberger},\ and\ \citenamefont
  {Krausz}}]{Cavalieri2007}%
  \BibitemOpen
  \bibfield  {author} {\bibinfo {author} {\bibfnamefont {A.}~\bibnamefont
  {Cavalieri}}, \bibinfo {author} {\bibfnamefont {N.}~\bibnamefont
  {M{\"u}ller}}, \bibinfo {author} {\bibfnamefont {T.}~\bibnamefont {Uphues}},
  \bibinfo {author} {\bibfnamefont {S.}~\bibnamefont {Thiele}}, \bibinfo
  {author} {\bibfnamefont {M.}~\bibnamefont {Schultze}}, \bibinfo {author}
  {\bibfnamefont {G.}~\bibnamefont {Sansone}}, \bibinfo {author} {\bibfnamefont
  {O.~D.}\ \bibnamefont {M{\"u}cke}}, \bibinfo {author} {\bibfnamefont
  {S.}~\bibnamefont {Nolte}}, \bibinfo {author} {\bibfnamefont
  {S.}~\bibnamefont {Stamm}}, \bibinfo {author} {\bibfnamefont
  {G.}~\bibnamefont {Pr{\"u}mper}}, \bibinfo {author} {\bibfnamefont
  {J.}~\bibnamefont {Rossbach}}, \bibinfo {author} {\bibfnamefont {C.~D.}\
  \bibnamefont {Schr{\"o}ter}}, \bibinfo {author} {\bibfnamefont
  {T.}~\bibnamefont {Popmintchev}}, \bibinfo {author} {\bibfnamefont
  {S.}~\bibnamefont {H{\"a}drich}}, \bibinfo {author} {\bibfnamefont {M.~F.}\
  \bibnamefont {Kling}}, \bibinfo {author} {\bibfnamefont {M.}~\bibnamefont
  {Locher}}, \bibinfo {author} {\bibfnamefont {J.}~\bibnamefont {Schreiber}},
  \bibinfo {author} {\bibfnamefont {A.}~\bibnamefont {Pietsch}}, \bibinfo
  {author} {\bibfnamefont {D.}~\bibnamefont {Azoury}}, \bibinfo {author}
  {\bibfnamefont {M.}~\bibnamefont {Fischer}}, \bibinfo {author} {\bibfnamefont
  {A.}~\bibnamefont {Monmayrant}}, \bibinfo {author} {\bibfnamefont
  {R.}~\bibnamefont {Kienberger}},\ and\ \bibinfo {author} {\bibfnamefont
  {F.}~\bibnamefont {Krausz}},\ }\bibfield  {title} {\bibinfo {title}
  {Attosecond spectroscopy in condensed matter},\ }\href
  {https://doi.org/10.1038/nature06229} {\bibfield  {journal} {\bibinfo
  {journal} {Nature}\ }\textbf {\bibinfo {volume} {449}},\ \bibinfo {pages}
  {1029} (\bibinfo {year} {2007})}\BibitemShut {NoStop}%
\bibitem [{\citenamefont {Krausz}\ and\ \citenamefont
  {Ivanov}(2009)}]{RevModPhys.81.163}%
  \BibitemOpen
  \bibfield  {author} {\bibinfo {author} {\bibfnamefont {F.}~\bibnamefont
  {Krausz}}\ and\ \bibinfo {author} {\bibfnamefont {M.}~\bibnamefont
  {Ivanov}},\ }\bibfield  {title} {\bibinfo {title} {Attosecond physics},\
  }\href {https://doi.org/10.1103/RevModPhys.81.163} {\bibfield  {journal}
  {\bibinfo  {journal} {Rev. Mod. Phys.}\ }\textbf {\bibinfo {volume} {81}},\
  \bibinfo {pages} {163} (\bibinfo {year} {2009})}\BibitemShut {NoStop}%
\bibitem [{\citenamefont {Lewenstein}\ \emph {et~al.}(1994)\citenamefont
  {Lewenstein}, \citenamefont {Balcou}, \citenamefont {Ivanov}, \citenamefont
  {L'Huillier},\ and\ \citenamefont {Corkum}}]{PhysRevA.49.2117}%
  \BibitemOpen
  \bibfield  {author} {\bibinfo {author} {\bibfnamefont {M.}~\bibnamefont
  {Lewenstein}}, \bibinfo {author} {\bibfnamefont {P.}~\bibnamefont {Balcou}},
  \bibinfo {author} {\bibfnamefont {M.~Y.}\ \bibnamefont {Ivanov}}, \bibinfo
  {author} {\bibfnamefont {A.}~\bibnamefont {L'Huillier}},\ and\ \bibinfo
  {author} {\bibfnamefont {P.~B.}\ \bibnamefont {Corkum}},\ }\bibfield  {title}
  {\bibinfo {title} {Theory of high-harmonic generation by low-frequency laser
  fields},\ }\href {https://doi.org/10.1103/PhysRevA.49.2117} {\bibfield
  {journal} {\bibinfo  {journal} {Phys. Rev. A}\ }\textbf {\bibinfo {volume}
  {49}},\ \bibinfo {pages} {2117} (\bibinfo {year} {1994})}\BibitemShut
  {NoStop}%
\bibitem [{\citenamefont {Corkum}(1993)}]{PhysRevLett.71.1994}%
  \BibitemOpen
  \bibfield  {author} {\bibinfo {author} {\bibfnamefont {P.~B.}\ \bibnamefont
  {Corkum}},\ }\bibfield  {title} {\bibinfo {title} {Plasma perspective on
  strong field multiphoton ionization},\ }\href
  {https://doi.org/10.1103/PhysRevLett.71.1994} {\bibfield  {journal} {\bibinfo
   {journal} {Phys. Rev. Lett.}\ }\textbf {\bibinfo {volume} {71}},\ \bibinfo
  {pages} {1994} (\bibinfo {year} {1993})}\BibitemShut {NoStop}%
\bibitem [{\citenamefont {Ishikawa}\ and\ \citenamefont {Sato}(2015)}]{Ishika}%
  \BibitemOpen
  \bibfield  {author} {\bibinfo {author} {\bibfnamefont {K.~L.}\ \bibnamefont
  {Ishikawa}}\ and\ \bibinfo {author} {\bibfnamefont {T.}~\bibnamefont
  {Sato}},\ }\bibfield  {title} {\bibinfo {title} {A review on ab initio
  approaches for multielectron dynamics},\ }\href
  {https://doi.org/10.1109/JSTQE.2015.2438827} {\bibfield  {journal} {\bibinfo
  {journal} {IEEE Journal of Selected Topics in Quantum Electronics}\ }\textbf
  {\bibinfo {volume} {21}},\ \bibinfo {pages} {1} (\bibinfo {year}
  {2015})}\BibitemShut {NoStop}%
\bibitem [{\citenamefont {Gallmann}\ \emph {et~al.}(2012)\citenamefont
  {Gallmann}, \citenamefont {Cirelli},\ and\ \citenamefont {Keller}}]{KO}%
  \BibitemOpen
  \bibfield  {author} {\bibinfo {author} {\bibfnamefont {L.}~\bibnamefont
  {Gallmann}}, \bibinfo {author} {\bibfnamefont {C.}~\bibnamefont {Cirelli}},\
  and\ \bibinfo {author} {\bibfnamefont {U.}~\bibnamefont {Keller}},\
  }\bibfield  {title} {\bibinfo {title} {Attosecond science: Recent highlights
  and future trends},\ }\href
  {https://doi.org/10.1146/annurev-physchem-032511-143702} {\bibfield
  {journal} {\bibinfo  {journal} {Annual Review of Physical Chemistry}\
  }\textbf {\bibinfo {volume} {63}},\ \bibinfo {pages} {447} (\bibinfo {year}
  {2012})},\ \bibinfo {note} {pMID: 22404594},\ \Eprint
  {https://arxiv.org/abs/https://doi.org/10.1146/annurev-physchem-032511-143702}
  {https://doi.org/10.1146/annurev-physchem-032511-143702} \BibitemShut
  {NoStop}%
\bibitem [{\citenamefont {Agostini}\ and\ \citenamefont
  {DiMauro}(2004)}]{Agostini_2004}%
  \BibitemOpen
  \bibfield  {author} {\bibinfo {author} {\bibfnamefont {P.}~\bibnamefont
  {Agostini}}\ and\ \bibinfo {author} {\bibfnamefont {L.~F.}\ \bibnamefont
  {DiMauro}},\ }\bibfield  {title} {\bibinfo {title} {The physics of attosecond
  light pulses},\ }\href {https://doi.org/10.1088/0034-4885/67/6/R01}
  {\bibfield  {journal} {\bibinfo  {journal} {Reports on Progress in Physics}\
  }\textbf {\bibinfo {volume} {67}},\ \bibinfo {pages} {813} (\bibinfo {year}
  {2004})}\BibitemShut {NoStop}%
\bibitem [{\citenamefont {Constant}\ \emph {et~al.}(1999)\citenamefont
  {Constant}, \citenamefont {Garzella}, \citenamefont {Breger}, \citenamefont
  {Mével}, \citenamefont {Dorrer}, \citenamefont {Blanc}, \citenamefont
  {Salin},\ and\ \citenamefont {Agostini}}]{Constant_1999}%
  \BibitemOpen
  \bibfield  {author} {\bibinfo {author} {\bibfnamefont {E.}~\bibnamefont
  {Constant}}, \bibinfo {author} {\bibfnamefont {D.}~\bibnamefont {Garzella}},
  \bibinfo {author} {\bibfnamefont {P.}~\bibnamefont {Breger}}, \bibinfo
  {author} {\bibfnamefont {E.}~\bibnamefont {Mével}}, \bibinfo {author}
  {\bibfnamefont {C.}~\bibnamefont {Dorrer}}, \bibinfo {author} {\bibfnamefont
  {C.~L.}\ \bibnamefont {Blanc}}, \bibinfo {author} {\bibfnamefont
  {F.}~\bibnamefont {Salin}},\ and\ \bibinfo {author} {\bibfnamefont
  {P.}~\bibnamefont {Agostini}},\ }\bibfield  {title} {\bibinfo {title}
  {Optimizing high harmonic generation in absorbing gases: Model and
  experiment},\ }\href {https://doi.org/10.1103/PhysRevLett.82.1668} {\bibfield
   {journal} {\bibinfo  {journal} {Physical Review Letters}\ }\textbf {\bibinfo
  {volume} {82}},\ \bibinfo {pages} {1668} (\bibinfo {year}
  {1999})}\BibitemShut {NoStop}%
\bibitem [{\citenamefont {Vampa}\ \emph {et~al.}(2015)\citenamefont {Vampa},
  \citenamefont {Hammond}, \citenamefont {Thir\'e}, \citenamefont {Schmidt},
  \citenamefont {L\'egar\'e}, \citenamefont {McDonald}, \citenamefont {Brabec},
  \citenamefont {Klug},\ and\ \citenamefont {Corkum}}]{PhysRevLett.115.193603}%
  \BibitemOpen
  \bibfield  {author} {\bibinfo {author} {\bibfnamefont {G.}~\bibnamefont
  {Vampa}}, \bibinfo {author} {\bibfnamefont {T.~J.}\ \bibnamefont {Hammond}},
  \bibinfo {author} {\bibfnamefont {N.}~\bibnamefont {Thir\'e}}, \bibinfo
  {author} {\bibfnamefont {B.~E.}\ \bibnamefont {Schmidt}}, \bibinfo {author}
  {\bibfnamefont {F.}~\bibnamefont {L\'egar\'e}}, \bibinfo {author}
  {\bibfnamefont {C.~R.}\ \bibnamefont {McDonald}}, \bibinfo {author}
  {\bibfnamefont {T.}~\bibnamefont {Brabec}}, \bibinfo {author} {\bibfnamefont
  {D.~D.}\ \bibnamefont {Klug}},\ and\ \bibinfo {author} {\bibfnamefont
  {P.~B.}\ \bibnamefont {Corkum}},\ }\bibfield  {title} {\bibinfo {title}
  {All-optical reconstruction of crystal band structure},\ }\href
  {https://doi.org/10.1103/PhysRevLett.115.193603} {\bibfield  {journal}
  {\bibinfo  {journal} {Phys. Rev. Lett.}\ }\textbf {\bibinfo {volume} {115}},\
  \bibinfo {pages} {193603} (\bibinfo {year} {2015})}\BibitemShut {NoStop}%
\bibitem [{\citenamefont {Ghimire}\ \emph {et~al.}(2011)\citenamefont
  {Ghimire}, \citenamefont {DiChiara}, \citenamefont {Sistrunk}, \citenamefont
  {Heinz},\ and\ \citenamefont {Leone}}]{Ghimire2011}%
  \BibitemOpen
  \bibfield  {author} {\bibinfo {author} {\bibfnamefont {S.}~\bibnamefont
  {Ghimire}}, \bibinfo {author} {\bibfnamefont {A.}~\bibnamefont {DiChiara}},
  \bibinfo {author} {\bibfnamefont {E.}~\bibnamefont {Sistrunk}}, \bibinfo
  {author} {\bibfnamefont {T.}~\bibnamefont {Heinz}},\ and\ \bibinfo {author}
  {\bibfnamefont {S.}~\bibnamefont {Leone}},\ }\bibfield  {title} {\bibinfo
  {title} {Observation of high-order harmonic generation in a bulk crystal},\
  }\href {https://doi.org/10.1038/nphys1847} {\bibfield  {journal} {\bibinfo
  {journal} {Nature Physics}\ }\textbf {\bibinfo {volume} {7}},\ \bibinfo
  {pages} {138} (\bibinfo {year} {2011})}\BibitemShut {NoStop}%
\bibitem [{\citenamefont {Murakami}\ \emph {et~al.}(2018)\citenamefont
  {Murakami}, \citenamefont {Eckstein},\ and\ \citenamefont {Werner}}]{IP1}%
  \BibitemOpen
  \bibfield  {author} {\bibinfo {author} {\bibfnamefont {Y.}~\bibnamefont
  {Murakami}}, \bibinfo {author} {\bibfnamefont {M.}~\bibnamefont {Eckstein}},\
  and\ \bibinfo {author} {\bibfnamefont {P.}~\bibnamefont {Werner}},\
  }\bibfield  {title} {\bibinfo {title} {High-harmonic generation in mott
  insulators},\ }\href {https://doi.org/10.1103/PhysRevLett.121.057405}
  {\bibfield  {journal} {\bibinfo  {journal} {Phys. Rev. Lett.}\ }\textbf
  {\bibinfo {volume} {121}},\ \bibinfo {pages} {057405} (\bibinfo {year}
  {2018})}\BibitemShut {NoStop}%
\bibitem [{\citenamefont {Tancogne-Dejean}\ \emph {et~al.}(2018)\citenamefont
  {Tancogne-Dejean}, \citenamefont {Sentef},\ and\ \citenamefont
  {Rubio}}]{IP2}%
  \BibitemOpen
  \bibfield  {author} {\bibinfo {author} {\bibfnamefont {N.}~\bibnamefont
  {Tancogne-Dejean}}, \bibinfo {author} {\bibfnamefont {M.~A.}\ \bibnamefont
  {Sentef}},\ and\ \bibinfo {author} {\bibfnamefont {A.}~\bibnamefont
  {Rubio}},\ }\bibfield  {title} {\bibinfo {title} {Ultrafast modification of
  hubbard $u$ in a strongly correlated material: Ab initio high-harmonic
  generation in nio},\ }\href {https://doi.org/10.1103/PhysRevLett.121.097402}
  {\bibfield  {journal} {\bibinfo  {journal} {Phys. Rev. Lett.}\ }\textbf
  {\bibinfo {volume} {121}},\ \bibinfo {pages} {097402} (\bibinfo {year}
  {2018})}\BibitemShut {NoStop}%
\bibitem [{\citenamefont {Murakami}\ and\ \citenamefont {Werner}(2018)}]{IP3}%
  \BibitemOpen
  \bibfield  {author} {\bibinfo {author} {\bibfnamefont {Y.}~\bibnamefont
  {Murakami}}\ and\ \bibinfo {author} {\bibfnamefont {P.}~\bibnamefont
  {Werner}},\ }\bibfield  {title} {\bibinfo {title} {Nonequilibrium steady
  states of electric field driven mott insulators},\ }\href
  {https://doi.org/10.1103/PhysRevB.98.075102} {\bibfield  {journal} {\bibinfo
  {journal} {Phys. Rev. B}\ }\textbf {\bibinfo {volume} {98}},\ \bibinfo
  {pages} {075102} (\bibinfo {year} {2018})}\BibitemShut {NoStop}%
\bibitem [{\citenamefont {Silva}\ \emph {et~al.}(2018)\citenamefont {Silva},
  \citenamefont {Blinov}, \citenamefont {Rubtsov}, \citenamefont {Smirnova},\
  and\ \citenamefont {Ivanov}}]{IP4}%
  \BibitemOpen
  \bibfield  {author} {\bibinfo {author} {\bibfnamefont {R.}~\bibnamefont
  {Silva}}, \bibinfo {author} {\bibfnamefont {I.}~\bibnamefont {Blinov}},
  \bibinfo {author} {\bibfnamefont {A.}~\bibnamefont {Rubtsov}}, \bibinfo
  {author} {\bibfnamefont {O.}~\bibnamefont {Smirnova}},\ and\ \bibinfo
  {author} {\bibfnamefont {M.}~\bibnamefont {Ivanov}},\ }\bibfield  {title}
  {\bibinfo {title} {High-harmonic spectroscopy of ultrafast many-body dynamics
  in strongly correlated systems},\ }\href
  {https://doi.org/10.1038/s41566-018-0129-0} {\bibfield  {journal} {\bibinfo
  {journal} {Nature Photonics}\ }\textbf {\bibinfo {volume} {12}},\ \bibinfo
  {pages} {266} (\bibinfo {year} {2018})}\BibitemShut {NoStop}%
\bibitem [{\citenamefont {Roy}\ \emph {et~al.}(2020)\citenamefont {Roy},
  \citenamefont {Bera},\ and\ \citenamefont {Saha}}]{KS1}%
  \BibitemOpen
  \bibfield  {author} {\bibinfo {author} {\bibfnamefont {A.}~\bibnamefont
  {Roy}}, \bibinfo {author} {\bibfnamefont {S.}~\bibnamefont {Bera}},\ and\
  \bibinfo {author} {\bibfnamefont {K.}~\bibnamefont {Saha}},\ }\bibfield
  {title} {\bibinfo {title} {Nonlinear dynamical response of interacting bosons
  to synthetic electric field},\ }\href
  {https://doi.org/10.1103/PhysRevResearch.2.043133} {\bibfield  {journal}
  {\bibinfo  {journal} {Phys. Rev. Res.}\ }\textbf {\bibinfo {volume} {2}},\
  \bibinfo {pages} {043133} (\bibinfo {year} {2020})}\BibitemShut {NoStop}%
\bibitem [{\citenamefont {Dutta}\ \emph {et~al.}(2023)\citenamefont {Dutta},
  \citenamefont {Roy},\ and\ \citenamefont {Saha}}]{KS2}%
  \BibitemOpen
  \bibfield  {author} {\bibinfo {author} {\bibfnamefont {D.}~\bibnamefont
  {Dutta}}, \bibinfo {author} {\bibfnamefont {A.}~\bibnamefont {Roy}},\ and\
  \bibinfo {author} {\bibfnamefont {K.}~\bibnamefont {Saha}},\ }\bibfield
  {title} {\bibinfo {title} {Nonlinear response of interacting bosons in a
  quasiperiodic potential},\ }\href
  {https://doi.org/10.1103/PhysRevB.107.035120} {\bibfield  {journal} {\bibinfo
   {journal} {Phys. Rev. B}\ }\textbf {\bibinfo {volume} {107}},\ \bibinfo
  {pages} {035120} (\bibinfo {year} {2023})}\BibitemShut {NoStop}%
\bibitem [{\citenamefont {Chen}\ and\ \citenamefont
  {Bian}(2023)}]{PhysRevA.107.043111}%
  \BibitemOpen
  \bibfield  {author} {\bibinfo {author} {\bibfnamefont {J.-X.}\ \bibnamefont
  {Chen}}\ and\ \bibinfo {author} {\bibfnamefont {X.-B.}\ \bibnamefont
  {Bian}},\ }\bibfield  {title} {\bibinfo {title} {Theoretical analysis of
  high-order harmonic generation in liquids by a semiclassical method},\ }\href
  {https://doi.org/10.1103/PhysRevA.107.043111} {\bibfield  {journal} {\bibinfo
   {journal} {Phys. Rev. A}\ }\textbf {\bibinfo {volume} {107}},\ \bibinfo
  {pages} {043111} (\bibinfo {year} {2023})}\BibitemShut {NoStop}%
\bibitem [{\citenamefont {Ngoko~Djiokap}\ and\ \citenamefont
  {Starace}(2013)}]{Be}%
  \BibitemOpen
  \bibfield  {author} {\bibinfo {author} {\bibfnamefont {J.~M.}\ \bibnamefont
  {Ngoko~Djiokap}}\ and\ \bibinfo {author} {\bibfnamefont {A.~F.}\ \bibnamefont
  {Starace}},\ }\bibfield  {title} {\bibinfo {title} {Resonant enhancement of
  the harmonic-generation spectrum of beryllium},\ }\href
  {https://doi.org/10.1103/PhysRevA.88.053412} {\bibfield  {journal} {\bibinfo
  {journal} {Phys. Rev. A}\ }\textbf {\bibinfo {volume} {88}},\ \bibinfo
  {pages} {053412} (\bibinfo {year} {2013})}\BibitemShut {NoStop}%
\bibitem [{\citenamefont {Plummer}\ and\ \citenamefont
  {Noble}(2002)}]{Plummer_2002}%
  \BibitemOpen
  \bibfield  {author} {\bibinfo {author} {\bibfnamefont {M.}~\bibnamefont
  {Plummer}}\ and\ \bibinfo {author} {\bibfnamefont {C.~J.}\ \bibnamefont
  {Noble}},\ }\bibfield  {title} {\bibinfo {title} {Resonant enhancement of
  harmonic generation in argon at 248 nm},\ }\href
  {https://doi.org/10.1088/0953-4075/35/2/101} {\bibfield  {journal} {\bibinfo
  {journal} {Journal of Physics B: Atomic, Molecular and Optical Physics}\
  }\textbf {\bibinfo {volume} {35}},\ \bibinfo {pages} {L51} (\bibinfo {year}
  {2002})}\BibitemShut {NoStop}%
\bibitem [{\citenamefont {Ivanov}\ and\ \citenamefont {Kheifets}(2008)}]{Iva}%
  \BibitemOpen
  \bibfield  {author} {\bibinfo {author} {\bibfnamefont {I.~A.}\ \bibnamefont
  {Ivanov}}\ and\ \bibinfo {author} {\bibfnamefont {A.~S.}\ \bibnamefont
  {Kheifets}},\ }\bibfield  {title} {\bibinfo {title} {Resonant enhancement of
  generation of harmonics},\ }\href
  {https://doi.org/10.1103/PhysRevA.78.053406} {\bibfield  {journal} {\bibinfo
  {journal} {Phys. Rev. A}\ }\textbf {\bibinfo {volume} {78}},\ \bibinfo
  {pages} {053406} (\bibinfo {year} {2008})}\BibitemShut {NoStop}%
\bibitem [{\citenamefont {Shaaran}\ \emph {et~al.}(2012)\citenamefont
  {Shaaran}, \citenamefont {Ciappina},\ and\ \citenamefont {Lewenstein}}]{QU}%
  \BibitemOpen
  \bibfield  {author} {\bibinfo {author} {\bibfnamefont {T.}~\bibnamefont
  {Shaaran}}, \bibinfo {author} {\bibfnamefont {M.~F.}\ \bibnamefont
  {Ciappina}},\ and\ \bibinfo {author} {\bibfnamefont {M.}~\bibnamefont
  {Lewenstein}},\ }\bibfield  {title} {\bibinfo {title} {Quantum-orbit analysis
  of high-order-harmonic generation by resonant plasmon field enhancement},\
  }\href {https://doi.org/10.1103/PhysRevA.86.023408} {\bibfield  {journal}
  {\bibinfo  {journal} {Phys. Rev. A}\ }\textbf {\bibinfo {volume} {86}},\
  \bibinfo {pages} {023408} (\bibinfo {year} {2012})}\BibitemShut {NoStop}%
\bibitem [{\citenamefont {Wang}\ \emph {et~al.}(2017)\citenamefont {Wang},
  \citenamefont {He}, \citenamefont {Wang}, \citenamefont {Yuan}, \citenamefont
  {Zhu}, \citenamefont {Lan},\ and\ \citenamefont {Lu}}]{Wang:17}%
  \BibitemOpen
  \bibfield  {author} {\bibinfo {author} {\bibfnamefont {B.}~\bibnamefont
  {Wang}}, \bibinfo {author} {\bibfnamefont {L.}~\bibnamefont {He}}, \bibinfo
  {author} {\bibfnamefont {F.}~\bibnamefont {Wang}}, \bibinfo {author}
  {\bibfnamefont {H.}~\bibnamefont {Yuan}}, \bibinfo {author} {\bibfnamefont
  {X.}~\bibnamefont {Zhu}}, \bibinfo {author} {\bibfnamefont {P.}~\bibnamefont
  {Lan}},\ and\ \bibinfo {author} {\bibfnamefont {P.}~\bibnamefont {Lu}},\
  }\bibfield  {title} {\bibinfo {title} {Resonance enhanced high-order harmonic
  generation in h2$+$ by two sequential laser pulses},\ }\href
  {https://doi.org/10.1364/OE.25.017777} {\bibfield  {journal} {\bibinfo
  {journal} {Opt. Express}\ }\textbf {\bibinfo {volume} {25}},\ \bibinfo
  {pages} {17777} (\bibinfo {year} {2017})}\BibitemShut {NoStop}%
\bibitem [{\citenamefont {Yoshikawa}\ \emph {et~al.}(2019)\citenamefont
  {Yoshikawa}, \citenamefont {Nagai}, \citenamefont {Uchida}, \citenamefont
  {Takaguchi}, \citenamefont {Sasaki}, \citenamefont {Miyata},\ and\
  \citenamefont {Tanaka}}]{Yoshikawa2019}%
  \BibitemOpen
  \bibfield  {author} {\bibinfo {author} {\bibfnamefont {N.}~\bibnamefont
  {Yoshikawa}}, \bibinfo {author} {\bibfnamefont {K.}~\bibnamefont {Nagai}},
  \bibinfo {author} {\bibfnamefont {K.}~\bibnamefont {Uchida}}, \bibinfo
  {author} {\bibfnamefont {Y.}~\bibnamefont {Takaguchi}}, \bibinfo {author}
  {\bibfnamefont {S.}~\bibnamefont {Sasaki}}, \bibinfo {author} {\bibfnamefont
  {Y.}~\bibnamefont {Miyata}},\ and\ \bibinfo {author} {\bibfnamefont
  {K.}~\bibnamefont {Tanaka}},\ }\bibfield  {title} {\bibinfo {title}
  {Interband resonant high-harmonic generation by valley polarized
  electron-hole pairs},\ }\href {https://doi.org/10.1038/s41467-019-11697-6}
  {\bibfield  {journal} {\bibinfo  {journal} {Nature Communications}\ }\textbf
  {\bibinfo {volume} {10}},\ \bibinfo {pages} {3709} (\bibinfo {year}
  {2019})}\BibitemShut {NoStop}%
\bibitem [{\citenamefont {Singh}\ \emph {et~al.}(2024)\citenamefont {Singh},
  \citenamefont {Fareed}, \citenamefont {Shirinabadi}, \citenamefont
  {Marcelino}, \citenamefont {Zhu}, \citenamefont {L\'egar\'e},\ and\
  \citenamefont {Ozaki}}]{SINGH2024100043}%
  \BibitemOpen
  \bibfield  {author} {\bibinfo {author} {\bibfnamefont {M.}~\bibnamefont
  {Singh}}, \bibinfo {author} {\bibfnamefont {M.~A.}\ \bibnamefont {Fareed}},
  \bibinfo {author} {\bibfnamefont {R.~G.}\ \bibnamefont {Shirinabadi}},
  \bibinfo {author} {\bibfnamefont {R.}~\bibnamefont {Marcelino}}, \bibinfo
  {author} {\bibfnamefont {F.}~\bibnamefont {Zhu}}, \bibinfo {author}
  {\bibfnamefont {F.}~\bibnamefont {L\'egar\'e}},\ and\ \bibinfo {author}
  {\bibfnamefont {T.}~\bibnamefont {Ozaki}},\ }\bibfield  {title} {\bibinfo
  {title} {Recent advances in high-order harmonic generation from laser-ablated
  plumes at the advanced laser light source laboratory},\ }\href
  {https://doi.org/https://doi.org/10.1016/j.fpp.2024.100043} {\bibfield
  {journal} {\bibinfo  {journal} {Fundamental Plasma Physics}\ ,\ \bibinfo
  {pages} {100043}} (\bibinfo {year} {2024})}\BibitemShut {NoStop}%
\bibitem [{\citenamefont {Bruner}\ \emph {et~al.}(2018)\citenamefont {Bruner},
  \citenamefont {Kr\"{u}ger}, \citenamefont {Pedatzur}, \citenamefont
  {Orenstein}, \citenamefont {Azoury},\ and\ \citenamefont
  {Dudovich}}]{Bruner:18}%
  \BibitemOpen
  \bibfield  {author} {\bibinfo {author} {\bibfnamefont {B.~D.}\ \bibnamefont
  {Bruner}}, \bibinfo {author} {\bibfnamefont {M.}~\bibnamefont {Kr\"{u}ger}},
  \bibinfo {author} {\bibfnamefont {O.}~\bibnamefont {Pedatzur}}, \bibinfo
  {author} {\bibfnamefont {G.}~\bibnamefont {Orenstein}}, \bibinfo {author}
  {\bibfnamefont {D.}~\bibnamefont {Azoury}},\ and\ \bibinfo {author}
  {\bibfnamefont {N.}~\bibnamefont {Dudovich}},\ }\bibfield  {title} {\bibinfo
  {title} {Robust enhancement of high harmonic generation via attosecond
  control of ionization},\ }\href {https://doi.org/10.1364/OE.26.009310}
  {\bibfield  {journal} {\bibinfo  {journal} {Opt. Express}\ }\textbf {\bibinfo
  {volume} {26}},\ \bibinfo {pages} {9310} (\bibinfo {year}
  {2018})}\BibitemShut {NoStop}%
\bibitem [{\citenamefont {Strelkov}(2016)}]{Atto_pulse}%
  \BibitemOpen
  \bibfield  {author} {\bibinfo {author} {\bibfnamefont {V.~V.}\ \bibnamefont
  {Strelkov}},\ }\bibfield  {title} {\bibinfo {title} {Attosecond-pulse
  production using resonantly enhanced high-order harmonics},\ }\href
  {https://doi.org/10.1103/PhysRevA.94.063420} {\bibfield  {journal} {\bibinfo
  {journal} {Phys. Rev. A}\ }\textbf {\bibinfo {volume} {94}},\ \bibinfo
  {pages} {063420} (\bibinfo {year} {2016})}\BibitemShut {NoStop}%
\bibitem [{\citenamefont {Shao}\ \emph {et~al.}(2022)\citenamefont {Shao},
  \citenamefont {Lu}, \citenamefont {Zhang}, \citenamefont {Yu}, \citenamefont
  {Tohyama},\ and\ \citenamefont {Lu}}]{QCP}%
  \BibitemOpen
  \bibfield  {author} {\bibinfo {author} {\bibfnamefont {C.}~\bibnamefont
  {Shao}}, \bibinfo {author} {\bibfnamefont {H.}~\bibnamefont {Lu}}, \bibinfo
  {author} {\bibfnamefont {X.}~\bibnamefont {Zhang}}, \bibinfo {author}
  {\bibfnamefont {C.}~\bibnamefont {Yu}}, \bibinfo {author} {\bibfnamefont
  {T.}~\bibnamefont {Tohyama}},\ and\ \bibinfo {author} {\bibfnamefont
  {R.}~\bibnamefont {Lu}},\ }\bibfield  {title} {\bibinfo {title}
  {High-harmonic generation approaching the quantum critical point of strongly
  correlated systems},\ }\href {https://doi.org/10.1103/PhysRevLett.128.047401}
  {\bibfield  {journal} {\bibinfo  {journal} {Phys. Rev. Lett.}\ }\textbf
  {\bibinfo {volume} {128}},\ \bibinfo {pages} {047401} (\bibinfo {year}
  {2022})}\BibitemShut {NoStop}%
\bibitem [{\citenamefont {Du}\ \emph {et~al.}(2016)\citenamefont {Du},
  \citenamefont {Guan}, \citenamefont {Zhou},\ and\ \citenamefont
  {Bian}}]{PhysRevA.94.023419}%
  \BibitemOpen
  \bibfield  {author} {\bibinfo {author} {\bibfnamefont {T.-Y.}\ \bibnamefont
  {Du}}, \bibinfo {author} {\bibfnamefont {Z.}~\bibnamefont {Guan}}, \bibinfo
  {author} {\bibfnamefont {X.-X.}\ \bibnamefont {Zhou}},\ and\ \bibinfo
  {author} {\bibfnamefont {X.-B.}\ \bibnamefont {Bian}},\ }\bibfield  {title}
  {\bibinfo {title} {Enhanced high-order harmonic generation from periodic
  potentials in inhomogeneous laser fields},\ }\href
  {https://doi.org/10.1103/PhysRevA.94.023419} {\bibfield  {journal} {\bibinfo
  {journal} {Phys. Rev. A}\ }\textbf {\bibinfo {volume} {94}},\ \bibinfo
  {pages} {023419} (\bibinfo {year} {2016})}\BibitemShut {NoStop}%
\bibitem [{\citenamefont {Fang}\ \emph {et~al.}(2024)\citenamefont {Fang},
  \citenamefont {Wang}, \citenamefont {Liu}, \citenamefont {Wang},
  \citenamefont {Cimmino}, \citenamefont {Wei}, \citenamefont {Bintz},
  \citenamefont {Parr}, \citenamefont {Kemp}, \citenamefont {Ni},\ and\
  \citenamefont {Yao}}]{Yao}%
  \BibitemOpen
  \bibfield  {author} {\bibinfo {author} {\bibfnamefont {F.}~\bibnamefont
  {Fang}}, \bibinfo {author} {\bibfnamefont {K.}~\bibnamefont {Wang}}, \bibinfo
  {author} {\bibfnamefont {V.~S.}\ \bibnamefont {Liu}}, \bibinfo {author}
  {\bibfnamefont {Y.}~\bibnamefont {Wang}}, \bibinfo {author} {\bibfnamefont
  {R.}~\bibnamefont {Cimmino}}, \bibinfo {author} {\bibfnamefont
  {J.}~\bibnamefont {Wei}}, \bibinfo {author} {\bibfnamefont {M.}~\bibnamefont
  {Bintz}}, \bibinfo {author} {\bibfnamefont {A.}~\bibnamefont {Parr}},
  \bibinfo {author} {\bibfnamefont {J.}~\bibnamefont {Kemp}}, \bibinfo {author}
  {\bibfnamefont {K.-K.}\ \bibnamefont {Ni}},\ and\ \bibinfo {author}
  {\bibfnamefont {N.~Y.}\ \bibnamefont {Yao}},\ }\href@noop {} {\bibinfo
  {title} {Probing critical phenomena in open quantum systems using atom
  arrays}} (\bibinfo {year} {2024}),\ \Eprint
  {https://arxiv.org/abs/2402.15376} {arXiv:2402.15376 [quant-ph]} \BibitemShut
  {NoStop}%
\bibitem [{\citenamefont {Song}\ \emph {et~al.}(2023)\citenamefont {Song},
  \citenamefont {Zhao}, \citenamefont {Janssen}, \citenamefont {Scherer},\ and\
  \citenamefont {Meng}}]{song2023deconfined}%
  \BibitemOpen
  \bibfield  {author} {\bibinfo {author} {\bibfnamefont {M.}~\bibnamefont
  {Song}}, \bibinfo {author} {\bibfnamefont {J.}~\bibnamefont {Zhao}}, \bibinfo
  {author} {\bibfnamefont {L.}~\bibnamefont {Janssen}}, \bibinfo {author}
  {\bibfnamefont {M.~M.}\ \bibnamefont {Scherer}},\ and\ \bibinfo {author}
  {\bibfnamefont {Z.~Y.}\ \bibnamefont {Meng}},\ }\href@noop {} {\bibinfo
  {title} {Deconfined quantum criticality lost}} (\bibinfo {year} {2023}),\
  \Eprint {https://arxiv.org/abs/2307.02547} {arXiv:2307.02547
  [cond-mat.str-el]} \BibitemShut {NoStop}%
\bibitem [{\citenamefont {Ejima}\ \emph {et~al.}(2014)\citenamefont {Ejima},
  \citenamefont {Lange},\ and\ \citenamefont {Fehske}}]{Ent}%
  \BibitemOpen
  \bibfield  {author} {\bibinfo {author} {\bibfnamefont {S.}~\bibnamefont
  {Ejima}}, \bibinfo {author} {\bibfnamefont {F.}~\bibnamefont {Lange}},\ and\
  \bibinfo {author} {\bibfnamefont {H.}~\bibnamefont {Fehske}},\ }\bibfield
  {title} {\bibinfo {title} {Spectral and entanglement properties of the
  bosonic haldane insulator},\ }\href
  {https://doi.org/10.1103/PhysRevLett.113.020401} {\bibfield  {journal}
  {\bibinfo  {journal} {Phys. Rev. Lett.}\ }\textbf {\bibinfo {volume} {113}},\
  \bibinfo {pages} {020401} (\bibinfo {year} {2014})}\BibitemShut {NoStop}%
\bibitem [{\citenamefont {Lin}\ \emph {et~al.}(2011)\citenamefont {Lin},
  \citenamefont {Compton}, \citenamefont {Jim{\'e}nez-Garc{\'i}a},
  \citenamefont {Phillips}, \citenamefont {Porto},\ and\ \citenamefont
  {Spielman}}]{synthetic}%
  \BibitemOpen
  \bibfield  {author} {\bibinfo {author} {\bibfnamefont {Y.-J.}\ \bibnamefont
  {Lin}}, \bibinfo {author} {\bibfnamefont {R.~L.}\ \bibnamefont {Compton}},
  \bibinfo {author} {\bibfnamefont {K.}~\bibnamefont {Jim{\'e}nez-Garc{\'i}a}},
  \bibinfo {author} {\bibfnamefont {W.~D.}\ \bibnamefont {Phillips}}, \bibinfo
  {author} {\bibfnamefont {J.~V.}\ \bibnamefont {Porto}},\ and\ \bibinfo
  {author} {\bibfnamefont {I.~B.}\ \bibnamefont {Spielman}},\ }\bibfield
  {title} {\bibinfo {title} {A synthetic electric force acting on neutral
  atoms},\ }\href {https://doi.org/10.1038/nphys1954} {\bibfield  {journal}
  {\bibinfo  {journal} {Nature Physics}\ }\textbf {\bibinfo {volume} {7}},\
  \bibinfo {pages} {531} (\bibinfo {year} {2011})}\BibitemShut {NoStop}%
\bibitem [{\citenamefont {Peierls}(1933)}]{Peierls1933}%
  \BibitemOpen
  \bibfield  {author} {\bibinfo {author} {\bibfnamefont {R.}~\bibnamefont
  {Peierls}},\ }\bibfield  {title} {\bibinfo {title} {Zur theorie des
  diamagnetismus von leitungselektronen},\ }\href
  {https://doi.org/10.1007/BF01342591} {\bibfield  {journal} {\bibinfo
  {journal} {Zeitschrift f{\"u}r Physik}\ }\textbf {\bibinfo {volume} {80}},\
  \bibinfo {pages} {763} (\bibinfo {year} {1933})}\BibitemShut {NoStop}%
\bibitem [{\citenamefont {Hauschild}\ and\ \citenamefont
  {Pollmann}(2018)}]{tenpy}%
  \BibitemOpen
  \bibfield  {author} {\bibinfo {author} {\bibfnamefont {J.}~\bibnamefont
  {Hauschild}}\ and\ \bibinfo {author} {\bibfnamefont {F.}~\bibnamefont
  {Pollmann}},\ }\bibfield  {title} {\bibinfo {title} {{Efficient numerical
  simulations with Tensor Networks: Tensor Network Python (TeNPy)}},\ }\href
  {https://doi.org/10.21468/SciPostPhysLectNotes.5} {\bibfield  {journal}
  {\bibinfo  {journal} {SciPost Phys. Lect. Notes}\ ,\ \bibinfo {pages} {5}}
  (\bibinfo {year} {2018})},\ \bibinfo {note} {code available from
  \url{https://github.com/tenpy/tenpy}},\ \Eprint
  {https://arxiv.org/abs/1805.00055} {arXiv:1805.00055} \BibitemShut {NoStop}%
\bibitem [{\citenamefont {Weinberg}\ and\ \citenamefont
  {Bukov}(2019)}]{Quspin}%
  \BibitemOpen
  \bibfield  {author} {\bibinfo {author} {\bibfnamefont {P.}~\bibnamefont
  {Weinberg}}\ and\ \bibinfo {author} {\bibfnamefont {M.}~\bibnamefont
  {Bukov}},\ }\bibfield  {title} {\bibinfo {title} {{QuSpin: a Python package
  for dynamics and exact diagonalisation of quantum many body systems. Part II:
  bosons, fermions and higher spins}},\ }\href
  {https://doi.org/10.21468/SciPostPhys.7.2.020} {\bibfield  {journal}
  {\bibinfo  {journal} {SciPost Phys.}\ }\textbf {\bibinfo {volume} {7}},\
  \bibinfo {pages} {020} (\bibinfo {year} {2019})}\BibitemShut {NoStop}%
\bibitem [{\citenamefont {Fishman}\ \emph {et~al.}(2022)\citenamefont
  {Fishman}, \citenamefont {White},\ and\ \citenamefont
  {Stoudenmire}}]{itensor}%
  \BibitemOpen
  \bibfield  {author} {\bibinfo {author} {\bibfnamefont {M.}~\bibnamefont
  {Fishman}}, \bibinfo {author} {\bibfnamefont {S.~R.}\ \bibnamefont {White}},\
  and\ \bibinfo {author} {\bibfnamefont {E.~M.}\ \bibnamefont {Stoudenmire}},\
  }\bibfield  {title} {\bibinfo {title} {{The ITensor Software Library for
  Tensor Network Calculations}},\ }\href
  {https://doi.org/10.21468/SciPostPhysCodeb.4} {\bibfield  {journal} {\bibinfo
   {journal} {SciPost Phys. Codebases}\ ,\ \bibinfo {pages} {4}} (\bibinfo
  {year} {2022})}\BibitemShut {NoStop}%
\bibitem [{\citenamefont {Gaarde}\ \emph {et~al.}(2008)\citenamefont {Gaarde},
  \citenamefont {Tate},\ and\ \citenamefont {Schafer}}]{Gaarde_2008}%
  \BibitemOpen
  \bibfield  {author} {\bibinfo {author} {\bibfnamefont {M.~B.}\ \bibnamefont
  {Gaarde}}, \bibinfo {author} {\bibfnamefont {J.~L.}\ \bibnamefont {Tate}},\
  and\ \bibinfo {author} {\bibfnamefont {K.~J.}\ \bibnamefont {Schafer}},\
  }\bibfield  {title} {\bibinfo {title} {Macroscopic aspects of attosecond
  pulse generation},\ }\href {https://doi.org/10.1088/0953-4075/41/13/132001}
  {\bibfield  {journal} {\bibinfo  {journal} {Journal of Physics B: Atomic,
  Molecular and Optical Physics}\ }\textbf {\bibinfo {volume} {41}},\ \bibinfo
  {pages} {132001} (\bibinfo {year} {2008})}\BibitemShut {NoStop}%
\bibitem [{\citenamefont {Sturm}(1993)}]{Sturm}%
  \BibitemOpen
  \bibfield  {author} {\bibinfo {author} {\bibfnamefont {K.}~\bibnamefont
  {Sturm}},\ }\bibfield  {title} {\bibinfo {title} {Dynamic structure factor:
  An introduction},\ }\href {https://doi.org/doi:10.1515/zna-1993-1-244}
  {\bibfield  {journal} {\bibinfo  {journal} {Zeitschrift für Naturforschung
  A}\ }\textbf {\bibinfo {volume} {48}},\ \bibinfo {pages} {233} (\bibinfo
  {year} {1993})}\BibitemShut {NoStop}%
\bibitem [{\citenamefont {Steinhauer}\ \emph {et~al.}(2002)\citenamefont
  {Steinhauer}, \citenamefont {Ozeri}, \citenamefont {Katz},\ and\
  \citenamefont {Davidson}}]{David}%
  \BibitemOpen
  \bibfield  {author} {\bibinfo {author} {\bibfnamefont {J.}~\bibnamefont
  {Steinhauer}}, \bibinfo {author} {\bibfnamefont {R.}~\bibnamefont {Ozeri}},
  \bibinfo {author} {\bibfnamefont {N.}~\bibnamefont {Katz}},\ and\ \bibinfo
  {author} {\bibfnamefont {N.}~\bibnamefont {Davidson}},\ }\bibfield  {title}
  {\bibinfo {title} {Excitation spectrum of a bose-einstein condensate},\
  }\href {https://doi.org/10.1103/PhysRevLett.88.120407} {\bibfield  {journal}
  {\bibinfo  {journal} {Phys. Rev. Lett.}\ }\textbf {\bibinfo {volume} {88}},\
  \bibinfo {pages} {120407} (\bibinfo {year} {2002})}\BibitemShut {NoStop}%
\bibitem [{\citenamefont {Ernst}\ \emph {et~al.}(2010)\citenamefont {Ernst},
  \citenamefont {G{\"o}tze}, \citenamefont {Krauser}, \citenamefont {Pyka},
  \citenamefont {L{\"u}hmann}, \citenamefont {Pfannkuche},\ and\ \citenamefont
  {Sengstock}}]{Bragg1}%
  \BibitemOpen
  \bibfield  {author} {\bibinfo {author} {\bibfnamefont {P.~T.}\ \bibnamefont
  {Ernst}}, \bibinfo {author} {\bibfnamefont {S.}~\bibnamefont {G{\"o}tze}},
  \bibinfo {author} {\bibfnamefont {J.~S.}\ \bibnamefont {Krauser}}, \bibinfo
  {author} {\bibfnamefont {K.}~\bibnamefont {Pyka}}, \bibinfo {author}
  {\bibfnamefont {D.-S.}\ \bibnamefont {L{\"u}hmann}}, \bibinfo {author}
  {\bibfnamefont {D.}~\bibnamefont {Pfannkuche}},\ and\ \bibinfo {author}
  {\bibfnamefont {K.}~\bibnamefont {Sengstock}},\ }\bibfield  {title} {\bibinfo
  {title} {Probing superfluids in optical lattices by momentum-resolved bragg
  spectroscopy},\ }\href {https://doi.org/10.1038/nphys1476} {\bibfield
  {journal} {\bibinfo  {journal} {Nature Physics}\ }\textbf {\bibinfo {volume}
  {6}},\ \bibinfo {pages} {56} (\bibinfo {year} {2010})}\BibitemShut {NoStop}%
\bibitem [{\citenamefont {Golinelli}\ \emph {et~al.}(1993)\citenamefont
  {Golinelli}, \citenamefont {Jolicoeur},\ and\ \citenamefont
  {Lacaze}}]{OGolinelli_1993}%
  \BibitemOpen
  \bibfield  {author} {\bibinfo {author} {\bibfnamefont {O.}~\bibnamefont
  {Golinelli}}, \bibinfo {author} {\bibfnamefont {T.}~\bibnamefont
  {Jolicoeur}},\ and\ \bibinfo {author} {\bibfnamefont {R.}~\bibnamefont
  {Lacaze}},\ }\bibfield  {title} {\bibinfo {title} {Dynamical properties of a
  haldane-gap antiferromagnet},\ }\href
  {https://doi.org/10.1088/0953-8984/5/9/024} {\bibfield  {journal} {\bibinfo
  {journal} {Journal of Physics: Condensed Matter}\ }\textbf {\bibinfo {volume}
  {5}},\ \bibinfo {pages} {1399} (\bibinfo {year} {1993})}\BibitemShut
  {NoStop}%
\bibitem [{\citenamefont {Huber}\ \emph {et~al.}(2007)\citenamefont {Huber},
  \citenamefont {Altman}, \citenamefont {B\"uchler},\ and\ \citenamefont
  {Blatter}}]{PhysRevB.75.085106}%
  \BibitemOpen
  \bibfield  {author} {\bibinfo {author} {\bibfnamefont {S.~D.}\ \bibnamefont
  {Huber}}, \bibinfo {author} {\bibfnamefont {E.}~\bibnamefont {Altman}},
  \bibinfo {author} {\bibfnamefont {H.~P.}\ \bibnamefont {B\"uchler}},\ and\
  \bibinfo {author} {\bibfnamefont {G.}~\bibnamefont {Blatter}},\ }\bibfield
  {title} {\bibinfo {title} {Dynamical properties of ultracold bosons in an
  optical lattice},\ }\href {https://doi.org/10.1103/PhysRevB.75.085106}
  {\bibfield  {journal} {\bibinfo  {journal} {Phys. Rev. B}\ }\textbf {\bibinfo
  {volume} {75}},\ \bibinfo {pages} {085106} (\bibinfo {year}
  {2007})}\BibitemShut {NoStop}%
\bibitem [{\citenamefont {Dalla~Torre}\ \emph {et~al.}(2006)\citenamefont
  {Dalla~Torre}, \citenamefont {Berg},\ and\ \citenamefont
  {Altman}}]{PhysRevLett.97.260401}%
  \BibitemOpen
  \bibfield  {author} {\bibinfo {author} {\bibfnamefont {E.~G.}\ \bibnamefont
  {Dalla~Torre}}, \bibinfo {author} {\bibfnamefont {E.}~\bibnamefont {Berg}},\
  and\ \bibinfo {author} {\bibfnamefont {E.}~\bibnamefont {Altman}},\
  }\bibfield  {title} {\bibinfo {title} {Hidden order in 1d bose insulators},\
  }\href {https://doi.org/10.1103/PhysRevLett.97.260401} {\bibfield  {journal}
  {\bibinfo  {journal} {Phys. Rev. Lett.}\ }\textbf {\bibinfo {volume} {97}},\
  \bibinfo {pages} {260401} (\bibinfo {year} {2006})}\BibitemShut {NoStop}%
\bibitem [{\citenamefont {Altman}\ and\ \citenamefont
  {Auerbach}(2002)}]{PhysRevLett.89.250404}%
  \BibitemOpen
  \bibfield  {author} {\bibinfo {author} {\bibfnamefont {E.}~\bibnamefont
  {Altman}}\ and\ \bibinfo {author} {\bibfnamefont {A.}~\bibnamefont
  {Auerbach}},\ }\bibfield  {title} {\bibinfo {title} {Oscillating
  superfluidity of bosons in optical lattices},\ }\href
  {https://doi.org/10.1103/PhysRevLett.89.250404} {\bibfield  {journal}
  {\bibinfo  {journal} {Phys. Rev. Lett.}\ }\textbf {\bibinfo {volume} {89}},\
  \bibinfo {pages} {250404} (\bibinfo {year} {2002})}\BibitemShut {NoStop}%
\bibitem [{\citenamefont {Ejima}\ and\ \citenamefont {Fehske}(2015)}]{SpinEnt}%
  \BibitemOpen
  \bibfield  {author} {\bibinfo {author} {\bibfnamefont {S.}~\bibnamefont
  {Ejima}}\ and\ \bibinfo {author} {\bibfnamefont {H.}~\bibnamefont {Fehske}},\
  }\bibfield  {title} {\bibinfo {title} {Comparative density-matrix
  renormalization group study of symmetry-protected topological phases in
  spin-1 chain and bose-hubbard models},\ }\href
  {https://doi.org/10.1103/PhysRevB.91.045121} {\bibfield  {journal} {\bibinfo
  {journal} {Phys. Rev. B}\ }\textbf {\bibinfo {volume} {91}},\ \bibinfo
  {pages} {045121} (\bibinfo {year} {2015})}\BibitemShut {NoStop}%
\bibitem [{\citenamefont {Weckesser}\ \emph {et~al.}(2024)\citenamefont
  {Weckesser}, \citenamefont {Srakaew}, \citenamefont {Blatz}, \citenamefont
  {Wei}, \citenamefont {Adler}, \citenamefont {Agrawal}, \citenamefont
  {Bohrdt}, \citenamefont {Bloch},\ and\ \citenamefont
  {Zeiher}}]{weckesser2024realizationrydbergdressedextendedbose}%
  \BibitemOpen
  \bibfield  {author} {\bibinfo {author} {\bibfnamefont {P.}~\bibnamefont
  {Weckesser}}, \bibinfo {author} {\bibfnamefont {K.}~\bibnamefont {Srakaew}},
  \bibinfo {author} {\bibfnamefont {T.}~\bibnamefont {Blatz}}, \bibinfo
  {author} {\bibfnamefont {D.}~\bibnamefont {Wei}}, \bibinfo {author}
  {\bibfnamefont {D.}~\bibnamefont {Adler}}, \bibinfo {author} {\bibfnamefont
  {S.}~\bibnamefont {Agrawal}}, \bibinfo {author} {\bibfnamefont
  {A.}~\bibnamefont {Bohrdt}}, \bibinfo {author} {\bibfnamefont
  {I.}~\bibnamefont {Bloch}},\ and\ \bibinfo {author} {\bibfnamefont
  {J.}~\bibnamefont {Zeiher}},\ }\href {https://arxiv.org/abs/2405.20128}
  {\bibinfo {title} {Realization of a rydberg-dressed extended bose hubbard
  model}} (\bibinfo {year} {2024}),\ \Eprint {https://arxiv.org/abs/2405.20128}
  {arXiv:2405.20128 [cond-mat.quant-gas]} \BibitemShut {NoStop}%
\bibitem [{\citenamefont {Lagoin}\ \emph {et~al.}(2022)\citenamefont {Lagoin},
  \citenamefont {Bhattacharya}, \citenamefont {Grass}, \citenamefont
  {Chhajlany}, \citenamefont {Salamon}, \citenamefont {Baldwin}, \citenamefont
  {Pfeiffer}, \citenamefont {Lewenstein}, \citenamefont {Holzmann},\ and\
  \citenamefont {Dubin}}]{Lagoin2022}%
  \BibitemOpen
  \bibfield  {author} {\bibinfo {author} {\bibfnamefont {C.}~\bibnamefont
  {Lagoin}}, \bibinfo {author} {\bibfnamefont {U.}~\bibnamefont
  {Bhattacharya}}, \bibinfo {author} {\bibfnamefont {T.}~\bibnamefont {Grass}},
  \bibinfo {author} {\bibfnamefont {R.~W.}\ \bibnamefont {Chhajlany}}, \bibinfo
  {author} {\bibfnamefont {T.}~\bibnamefont {Salamon}}, \bibinfo {author}
  {\bibfnamefont {K.}~\bibnamefont {Baldwin}}, \bibinfo {author} {\bibfnamefont
  {L.}~\bibnamefont {Pfeiffer}}, \bibinfo {author} {\bibfnamefont
  {M.}~\bibnamefont {Lewenstein}}, \bibinfo {author} {\bibfnamefont
  {M.}~\bibnamefont {Holzmann}},\ and\ \bibinfo {author} {\bibfnamefont
  {F.}~\bibnamefont {Dubin}},\ }\bibfield  {title} {\bibinfo {title} {Extended
  bose--hubbard model with dipolar excitons},\ }\href
  {https://doi.org/10.1038/s41586-022-05123-z} {\bibfield  {journal} {\bibinfo
  {journal} {Nature}\ }\textbf {\bibinfo {volume} {609}},\ \bibinfo {pages}
  {485} (\bibinfo {year} {2022})}\BibitemShut {NoStop}%
\bibitem [{\citenamefont {Arg\"uello-Luengo}\ \emph {et~al.}(2024)\citenamefont
  {Arg\"uello-Luengo}, \citenamefont {Rivera-Dean}, \citenamefont {Stammer},
  \citenamefont {Maxwell}, \citenamefont {Weld}, \citenamefont {Ciappina},\
  and\ \citenamefont {Lewenstein}}]{Weld}%
  \BibitemOpen
  \bibfield  {author} {\bibinfo {author} {\bibfnamefont {J.}~\bibnamefont
  {Arg\"uello-Luengo}}, \bibinfo {author} {\bibfnamefont {J.}~\bibnamefont
  {Rivera-Dean}}, \bibinfo {author} {\bibfnamefont {P.}~\bibnamefont
  {Stammer}}, \bibinfo {author} {\bibfnamefont {A.~S.}\ \bibnamefont
  {Maxwell}}, \bibinfo {author} {\bibfnamefont {D.~M.}\ \bibnamefont {Weld}},
  \bibinfo {author} {\bibfnamefont {M.~F.}\ \bibnamefont {Ciappina}},\ and\
  \bibinfo {author} {\bibfnamefont {M.}~\bibnamefont {Lewenstein}},\ }\bibfield
   {title} {\bibinfo {title} {Analog simulation of high-harmonic generation in
  atoms},\ }\href {https://doi.org/10.1103/PRXQuantum.5.010328} {\bibfield
  {journal} {\bibinfo  {journal} {PRX Quantum}\ }\textbf {\bibinfo {volume}
  {5}},\ \bibinfo {pages} {010328} (\bibinfo {year} {2024})}\BibitemShut
  {NoStop}%
\end{thebibliography}%
\end{document}